\documentclass{article}

\usepackage[preprint]{neurips_2026}

\usepackage[utf8]{inputenc}
\usepackage[T1]{fontenc}
\usepackage{url}
\usepackage{booktabs}
\usepackage{amsfonts}
\usepackage{amssymb}
\usepackage{amsthm}
\usepackage{amsmath}
\usepackage{mathtools}
\usepackage{nicefrac}
\usepackage{microtype}
\usepackage{xcolor}
\usepackage{graphicx}
\usepackage[colorlinks=true,linkcolor=blue!50!black,citecolor=blue!50!black,urlcolor=blue!50!black]{hyperref}
\usepackage[capitalize,nameinlink]{cleveref}
\usepackage{enumitem}
\usepackage{listings}
\usepackage{caption}
\usepackage{subcaption}

\theoremstyle{definition}

\graphicspath{{figures/}}

\title{Estimated Dynamic Equilibrium Model: Supply and Demand as a Sample Path of a Stochastic Process}

\author{%
  Mikhail L. Arbuzov \\
  Independent Researcher \\
  \texttt{Mike.arbuzov54@gmail.com} \\
  \And
  Sisong Bei \\
  Independent Researcher \\
  \texttt{qurining@gmail.com} \\
  \And
  Alexey Shvets \\
  Palo Alto Networks \\
  \texttt{ashvets@paloaltonetworks.com} \\
}

\begin{document}

\maketitle

\begin{abstract}
We introduce the \emph{Estimated Dynamic Equilibrium Model} (EDEM), an agent-based 
framework that treats supply and demand as a coupled stochastic process 
driven by heterogeneous, noisy agent valuations. The model’s primary 
technical contribution is the identification of a generative mechanism 
for persistent disequilibrium: when market-clearing prices are 
sequentially sampled from the upper tail of noisy bid distributions 
and recycled as inputs for future valuations, expected prices drift 
upward despite strictly zero-mean estimation errors. We derive this 
order-statistic bias in closed form for i.i.d. uniform bids and use 
simulations to show that compounding this bias across epochs yields 
exponential price growth without requiring assumptions of investor 
optimism or irrationality. This framework extends Miller’s 
divergence-of-opinion theory to a dynamic setting, recovering 
Walrasian equilibrium and Miller’s static premium as limiting cases. 
Through controlled experiments and sensitivity analysis on a simulated 
real-estate neighborhood, we identify six distinct regimes—ranging from 
band-stability to runaway bubbles—emerging from a single agent ruleset. 
These results offer a potential explanation for the contradictory 
findings in the empirical divergence-of-opinion literature and suggest 
that machine-learning valuation algorithms may inadvertently amplify 
this inherent statistical bias.
\end{abstract}


\section{Introduction}
\label{sec:intro}

The Efficient Market Hypothesis (EMH) remains the dominant theory of
price formation. It predicts when prices reflect available information
but does not provide a generative microstructural account of when and
why they fail to: bubbles, crashes, persistent cross-sectional return
anomalies, and the stylised observation that real markets very rarely
sit at the textbook supply--demand fixed point. Decades of empirical
work have produced rival theories whose claims often contradict one
another, because abundant and noisy financial data can be mined to
support nearly any hypothesis \citep{moffitt2017}. To avoid
contributing yet another empirical controversy, this paper takes the
less traveled \emph{atheoretical} route: constructing an economy from
the ground up in a controlled environment and asking what behavior
emerges from a small number of common-sense rules.

The central technical claim of the paper is that two structural
features --- max-bid clearing and per-epoch feedback of clearing
prices into agent valuations --- generate persistent positive drift
in realised market prices even when individual agents' estimation
errors are exactly zero-mean. The mechanism is order-statistic bias:
each seller selects the maximum of multiple noisy bids, and the
maximum of $n$ zero-mean draws is, in expectation, strictly above
their mean. Feeding the realised maximum back into the next round's
anchor compounds the bias multiplicatively. No behavioural assumption
about investor optimism, risk aversion, or attention is required.
Bubbles, business cycles, persistent overshoots, and persistent
undershoots all follow from this single mechanism interacting with
heterogeneous time-on-market dynamics.

We use agent-based modelling \citep{wilensky1999} because it allows
this kind of structural mechanism to emerge from a few simple,
transparent rules. The setting is a real-estate market in a small
neighbourhood --- a market in which every participant must form a
personal estimate of an asset whose true value is fundamentally
uncertain. Real estate is a useful laboratory: prices are observable,
transactions are costly enough that na\"ive arbitrage arguments fail,
and divergence of opinion among buyers and sellers is the rule
rather than the exception.

The central object of study is the \emph{Estimated Dynamic Equilibrium
Model} (EDEM), in which supply and demand are not static schedules
but realisations of a stochastic process driven by heterogeneous,
error-prone agent valuations. EDEM is a dynamic extension of
\citet{miller1977}'s divergence-of-opinion theory, but it relaxes
Miller's static one-shot setting to allow valuations, the price
level, and the population of buyers and sellers to co-evolve over
time.

The paper makes three contributions:
\begin{enumerate}
\item \textbf{An order-statistic mechanism for bubbles without
biased agents} (\cref{sec:framework-asymmetry}): we derive in closed
form, for a clean special case, that the expected winning bid
exceeds the home value by $\sigma(n-1)/(n+1)$ when $n$ bidders
draw zero-mean errors with dispersion $\sigma$. Compounded across
epochs, this produces the exponential price drift of
\cref{sec:exp-bubble}.

\item \textbf{A unified framework} (\cref{sec:framework}) that nests
the classical Walrasian equilibrium and Miller's static premium as
limiting cases, and admits a family of out-of-equilibrium regimes
(business cycles, persistent overshoots and undershoots, bubbles,
constant transition) parameterised by estimation noise, balancer
strength, and time-on-market.

\item \textbf{Eight primary controlled experiments plus a 30-cell
sensitivity grid}
(\cref{sec:experiments}) implemented in the Python ABM framework
Mesa \citep{masad2015mesa}, replacing earlier
\citet{wilensky1999} NetLogo prototypes. The Mesa codebase is open
source and reproduces every figure in the paper from a clean
checkout.
\end{enumerate}

A practical implication is that empirical premium-vs-discount
findings depend strongly on the sample's position along the
underlying stochastic trajectory, which short-window studies cannot
identify. Two studies that find opposite signs may both be correct
characterisations of distinct epochs in the same process. A
secondary implication concerns valuation algorithms: a
machine-learning estimator trained on historical clearing prices
inherits the bid-selection asymmetry and re-deploys it as a
positively-biased point estimator, providing the market with a
coordinating signal that can amplify the very bubble the algorithm
was meant to price. \Cref{sec:disc-ml} discusses this in connection
with the 2021 wind-down of Zillow Offers \citep{zillow2021ibuying}.

The remainder of the paper is organised as follows. \Cref{sec:background}
reviews \citet{miller1977} and the divergence-of-opinion debate.
\Cref{sec:framework} presents the EDEM framework formally and derives
the order-statistic drift. \Cref{sec:implementation} describes the
Mesa implementation. \Cref{sec:experiments} reports the eight primary
experiments and their results. \Cref{sec:discussion} discusses
implications. \Cref{sec:extensions} sketches extensions and
\cref{sec:conclusion} concludes.

\section{Background: Divergence of Opinion}
\label{sec:background}

\subsection{Miller's static premium hypothesis}
\label{sec:bg-miller}

\citet{miller1977} proposed a deceptively simple model of how heterogeneous
beliefs determine an asset's market price. Suppose any single investor can
purchase only one share---perhaps because of limited funds---and there are
$N$ shares available. The shares will end up owned by the $N$ investors with
the highest valuations. The marginal investor, the one with the lowest
valuation among the holders, sets the price. As the divergence of opinion
about the asset's value widens, this marginal valuation rises, pushing the
price above the mean estimate of all potential investors.

The implication is striking: when short selling is restricted, an asset's
price reflects the optimism of a minority, not the average view of the market.
Greater disagreement among investors thus produces a \emph{premium}, and
because eventual returns must be earned against this elevated price, future
returns on high-disagreement assets should be lower. This is the
``divergence of opinion premium hypothesis''---or, equivalently, the
\emph{overvaluation hypothesis} \citep{doukas2006}.

Miller offered initial empirical support: stocks about which there is the
greatest divergence of opinion at issuance experience smaller subsequent price
appreciation than ``seasoned'' stocks over horizons of one to five years.
Although Miller's model was largely ignored for decades, its narrative remains
unusually compelling. As \citet{moffitt2017} puts it, ``although Miller's
model has largely been ignored, it has lost none of its relevance.''

\subsection{The premium--discount debate}
\label{sec:bg-debate}

The empirical literature that grew up around Miller's hypothesis is
remarkable for its inability to converge. \citet{doukas2006} provide a careful
overview of the half-century of contested findings, of which the broad
outlines are as follows.

The \emph{discount} camp argues that divergence of opinion proxies for risk:
the higher the disagreement, the riskier the asset, and therefore the lower
its price relative to fundamentals \citep{williams1977,mayshar1983,
varian1985,merton1987,epsteinwang1994}. A direct consequence is that future
returns on high-disagreement assets should be \emph{higher}, not lower.

The \emph{premium} camp---supporting Miller's original prediction---finds
positive associations between dispersion in analyst earnings forecasts and
contemporaneous prices, with low subsequent returns. \citet{doukas2006}
themselves report evidence consistent with the discount hypothesis after
controlling for analyst-related dispersion measures, framing the apparent
contradictions as artefacts of measurement rather than substantive
disagreement.

Decades on, the debate is unresolved. \citet{moffitt2017} summarises the
state of play as follows:
\begin{quote}
``Though we have over two centuries of financial market history and various
theories of price formation, today there exists no single theory acceptable
to market practitioners, yet rigorous enough to satisfy market theorists.
The only major theory that purports universality is the efficient market
theory, which fails to explain some of the most important market phenomena,
e.g.\ bubbles and crashes.''
\end{quote}

\subsection{What is missing: dynamics}
\label{sec:bg-dynamics}

A common feature of Miller's model and its successors is that they are
fundamentally \emph{static}. They specify a one-shot relationship between the
distribution of beliefs and the market-clearing price. Real markets, however,
are continuous processes in which today's price feeds back into tomorrow's
beliefs, and in which the population of active buyers and sellers responds to
recent prices with adjustment lags.

This paper argues that the static framing is responsible for much of the
empirical confusion. When supply and demand are sample paths of a stochastic
process rather than fixed schedules, the same underlying mechanism---
heterogeneous valuations with random error---can produce \emph{either} a
divergence-of-opinion premium \emph{or} a discount, depending on the
balance between estimation noise and the market's adjustment capacity.
\Cref{sec:framework} formalises this claim.

\citet{shiller2003} reviews how behavioural finance has progressively
encroached on EMH territory; the present work occupies a related niche,
treating price formation itself as the locus of behavioural noise rather
than treating noise as a deviation from a frictionless benchmark.

\section{The EDEM Framework}
\label{sec:framework}

This section formalises the Estimated Dynamic Equilibrium Model. We
state EDEM in its general form, then identify the Dynamic Equilibrium
(DE) model of \cref{sec:experiments-de} and Walrasian equilibrium as
limiting cases. The central technical claim is that two structural
features --- max-bid clearing and per-epoch feedback of clearing
prices into agent valuations --- generate persistent positive drift
in realised prices even when individual agents' estimation errors are
zero-mean. \Cref{sec:framework-asymmetry} derives this drift in a
clean order-statistic form.

\subsection{Primitive objects}
\label{sec:framework-primitives}

A market is staged on a finite torus $G$ of homes; each home
$h \in G$ has a fixed but unobservable \emph{fair value} $v^{*}(h)$ and
a \emph{market value} $v_{t}(h) \in \mathbb{R}_{>0}$ that evolves over
discrete time $t = 0, 1, 2, \dots$. The market is populated by two
agent classes:
\begin{itemize}[leftmargin=*]
\item \textbf{Sellers} $\mathcal{S}_{t} \subset G$, each occupying one
      home and posting an ask price $a_{t}(s)$;
\item \textbf{Buyers} $\mathcal{B}_{t}$, each free to walk the torus
      and place bids on sellers it lands with.
\end{itemize}
We let $Q_{s}(t) = |\mathcal{S}_{t}|$ and $Q_{d}(t) = |\mathcal{B}_{t}|$
denote the supply and demand quantities at time $t$; both are
state variables, not parameters.

\subsection{Heterogeneous estimation}
\label{sec:framework-estimation}

Each agent is endowed with a personal \emph{estimation function}.
For an agent $i$ valuing home $h$ at time $t$:
\begin{equation}
\label{eq:estimation}
E_{i,t}(h) \;=\; v_{t}(h) \cdot \bigl(1 + \varepsilon_{i,t,h}\bigr),
\qquad
\varepsilon_{i,t,h} \;\stackrel{\mathrm{iid}}{\sim}\; F_{i}(\sigma_{i}),
\end{equation}
where $F_{i}$ is a zero-mean error distribution with agent-specific
dispersion $\sigma_{i}$. The three-way index is essential: each
\emph{bid} that agent $i$ places --- on a different home, or even on
the same home at a different tick --- draws an independent realisation
of the error. Without per-bid independence, the buyer-side
order-statistic logic of Cond.~2 (\cref{sec:framework-cond2}) and
the per-epoch update of \cref{eq:value-update} would degenerate to
single-draw statistics. Concretely,
$\mathbb{E}[\varepsilon_{i,t,h}] = 0$ and
$\operatorname{Var}(\varepsilon_{i,t,h})$ is increasing in
$\sigma_{i}$. The DE specialisation
(\cref{sec:framework-de-special}) takes $F_{i}$ as the symmetric
uniform $\mathcal{U}[-\sigma_{i},\,+\sigma_{i}]$; in the broader
EDEM the error distribution is left unspecified beyond zero-mean and
the support condition $\Pr(\varepsilon_{i,t} > -1) = 1$ that keeps
\cref{eq:estimation} strictly positive.

\paragraph{Unit convention.} Throughout the formal model,
$\sigma_{i}$ is a dimensionless dispersion parameter expressed as a
\emph{decimal}; e.g.\ $\sigma_{i} = 0.05$ corresponds to a $\pm 5\%$
maximum estimation error. The parameter tables and figure captions
report $\sigma$ as a percentage for readability; the conversion is
implicit when these values appear in \cref{eq:estimation} and its
descendants.

The dispersion parameter $\sigma_{i}$ is the \emph{divergence of
opinion} that \citet{miller1977} treats statically. Allowing
$\sigma_{i}$ to vary across agents (estimation heterogeneity) and over
time (rising or falling market-wide divergence of opinion) is the key
generalisation that makes EDEM a dynamic theory of price formation.

\subsection{Per-tick interaction}
\label{sec:framework-interaction}

At each tick:
\begin{enumerate}[leftmargin=*,nosep]
\item Buyers move and post bids: a buyer $b$ that lands on a seller
      $s$'s home draws a bid $\beta_{b,s}^{(t)} = E_{b,t}(h(s))$ from
      \cref{eq:estimation} and presents it to $s$. Bids accumulate;
      the seller tracks the best bid received.
\item Sellers process bids: when its patience timer elapses, $s$
      offers to sign with the buyer holding the best bid. The buyer
      either commits (a sale clears) or declines (the bid is
      dropped); the commitment rule is the key behavioural primitive
      and is discussed below in \cref{sec:framework-cond2}.
\end{enumerate}

\subsection{Market-clearing functional}
\label{sec:framework-clearing}

The realised market price at time $t$ is the output of a clearing
algorithm $A$ that consumes the full market state. Let
\begin{equation}
\label{eq:state}
X_{t} \;=\;
\bigl(\mathcal{S}_{t},\,\mathcal{B}_{t},\,
       \{v_{t}(h)\}_{h \in G},\,
       \{\beta_{\bullet}^{(t)}\},\,
       \{a_{t}(s)\}_{s \in \mathcal{S}_{t}},\,
       \{\text{patience}_{t}(s)\},\,
       \mathcal{H}_{t}\bigr)
\end{equation}
denote the model's state at $t$ (agent populations, home values,
outstanding bids and asks, seller patience timers, and the history
$\mathcal{H}_{t}$ of completed sales). The realised market price is
\begin{equation}
\label{eq:price-functional}
p_{t} \;=\; A(X_{t};\,\theta),
\end{equation}
where $\theta$ collects structural parameters. The reduced state
variables $Q_{s}(t)$, $Q_{d}(t)$, and a cross-population dispersion
summary $\sigma$ are the principal --- but not the only --- quantities
that $A$ depends on; we use the abbreviated notation
$A(Q_{s}(t), Q_{d}(t), \sigma)$ when the remaining state is held
implicit. In our two specific instantiations $A$ is realised
differently:
\begin{itemize}[leftmargin=*,nosep]
\item In DE, $A$ is the rolling average of the last $W$ \emph{sale
      events} (\cref{eq:price-rolling}). Sales are discrete events
      and not every tick records one.
\item In the speculative-market EDEM variant, $A$ is implicit in the
      multiplicative update rule \cref{eq:value-update}; sales are
      not realised, and the market price is the average home value.
\end{itemize}

\subsection{Stochastic supply and demand schedules}
\label{sec:framework-schedules}

Crucially, $Q_{s}$ and $Q_{d}$ in \cref{eq:price-functional} are not
the deterministic schedules $Q_{s}(p)$ and $Q_{d}(p)$ of textbook
microeconomics. They are integer-valued random walks driven by the
price increment $\Delta p_{t-1} = p_{t-1} - p_{t-2}$ via two balancer
parameters $C_{b}, C_{s}$:
\begin{align}
\label{eq:Qs-update}
Q_{s}(t) &= Q_{s}(t-1) + Z^{s}_{t},
& \mathbb{E}\bigl[Z^{s}_{t}\bigr] &= C_{b}\,\operatorname{sgn}(\Delta p_{t-1}), \\
\label{eq:Qd-update}
Q_{d}(t) &= Q_{d}(t-1) - Z^{d}_{t},
& \mathbb{E}\bigl[Z^{d}_{t}\bigr] &= C_{s}\,\operatorname{sgn}(\Delta p_{t-1}).
\end{align}
The integer increments $Z^{s}_{t}, Z^{d}_{t}$ are required because
agent counts are integers; they realise possibly non-integer expected
counts via the deterministic-plus-Bernoulli rule of
\cref{sec:implementation-balancer}, with a population floor of one
agent per side. With $C_{b}, C_{s} > 0$ rising prices add sellers
and remove buyers (mean-reverting); with $C_{b}, C_{s} < 0$ rising
prices add buyers and remove sellers (trend-following); with
$C_{b} = C_{s} = 0$ the populations are frozen and the price evolves
under the bid-selection asymmetry alone. The simulations in
\cref{sec:experiments-edem} use $C_{s} = C_{b}$ and report the
single number $C_{b}$.

\Cref{eq:Qs-update,eq:Qd-update} describe the \emph{interior}
dynamics. A population floor of one agent per side
(\cref{sec:implementation-balancer}) truncates the random walk
whenever $\max(Q_{s}, Q_{d})$ would otherwise dictate killing the
last remaining agent on the depleted side; under that boundary the
expectation equations no longer hold literally. The one-vs-many
regime that emerges for strong $|C_{b}|$ in Run~7
(\cref{sec:exp-balancer-sweep}) is exactly this boundary state.

Together,
\cref{eq:estimation,eq:price-functional,eq:Qs-update,eq:Qd-update}
define a coupled stochastic process whose realisation is a sample
path. Equilibrium in the textbook sense corresponds to the steady
state $\Delta p_{t} = \Delta Q_{s}(t) = \Delta Q_{d}(t) = 0$ --- a
non-generic state that the process visits only fleetingly under most
parameter regimes, which is the central observation of the paper.

\subsection{Bid commitment (Cond.~2)}
\label{sec:framework-cond2}

A buyer $b$ offered to sign at price $\beta$ commits iff $\beta$ is
at or above the buyer's own running benchmark over its outstanding
bids:
\begin{equation}
\label{eq:cond2}
b \text{ commits} \quad\Longleftrightarrow\quad \beta \;\ge\; \frac{1}{|\mathcal{B}_{b}|}\sum_{\beta' \in \mathcal{B}_{b}} \beta',
\end{equation}
where $\mathcal{B}_{b}$ is the set of all bids $b$ has currently
outstanding (the offered $\beta$ included, so $|\mathcal{B}_{b}| \ge 1$
always). The intuition is order-statistic selection on the buyer's
side: each $\beta' \in \mathcal{B}_{b}$ is a noisy estimate of fair
value, and the bids $b$ has placed on different sellers reflect
different draws of $b$'s estimation error. Committing only when the
offered $\beta$ is at or above the buyer's running benchmark
restricts $b$'s actual purchase to the upper tail of its own bid
distribution --- the homes for which $b$'s estimate happened to
land on the optimistic side. This is the buyer-side mirror of the
seller-side max-bid selection of \cref{sec:framework-asymmetry}, and
it is the second of the two ingredients that produces the order-
statistic drift. The prior write-up of the model in
\citet{arbuzov2018de} stated this rule with the inequality reversed;
we restore the version that matches the simulation results actually
reported there (\cref{sec:implementation-cond2}).

\subsection{The order-statistic drift: a special-case mechanism}
\label{sec:framework-asymmetry}

The bubble mechanism that Runs~6--8 exhibit is a consequence of
order-statistic bias. We give a closed-form result for a tractable
special case below, then connect it empirically to the actual update
rule \cref{eq:value-update}.

\paragraph{Closed form for the seller-side maximum.}
Consider a seller $s$ that has received $n \ge 2$ bids
$\beta_{1}, \dots, \beta_{n}$ from buyers with iid zero-mean
estimation errors $\varepsilon_{i} \sim \mathcal{U}[-\sigma, +\sigma]$:
\begin{equation}
\label{eq:bid-decomp}
\beta_{i} \;=\; v_{t}(h(s)) \cdot (1 + \varepsilon_{i}).
\end{equation}
The seller-level winning bid is $\max_{i} \beta_{i}$; its expected
ratio to the home's current value is
\begin{equation}
\label{eq:ostat-bias}
\mathbb{E}\!\left[\frac{\max_{i}\beta_{i}}{v_{t}(h(s))}\right]
\;=\;
1 + \mathbb{E}\!\left[\max_{i}\varepsilon_{i}\right]
\;=\;
1 + \sigma\,\frac{n - 1}{n + 1},
\end{equation}
using the standard order-statistic identity for the maximum of $n$
iid uniform variates. The right-hand side strictly exceeds one for
all $n \ge 2$ and all $\sigma > 0$; the gap grows in both $n$
(more bidders per seller) and $\sigma$ (wider divergence of opinion).

\paragraph{What \cref{eq:ostat-bias} does and does not prove.}
\Cref{eq:ostat-bias} establishes that the \emph{seller-side
maximum} has expected ratio strictly above one. The per-epoch update
in \cref{eq:value-update} compounds a related but distinct quantity
$\bar{r}_{t}$: a buyer-side mean over each buyer's $\min_{s}$ winning
ratio. We have not derived a closed form for $\bar{r}_{t}$. Two
further care points apply when moving from the clean theorem to the
sample-path behaviour of \cref{fig:run6-bubble}:
\begin{itemize}[leftmargin=*,nosep]
\item \emph{Expected level vs.\ typical trajectory.} For an iid
      multiplicative process $v_{T} = v_{0}\prod_{t}r_{t}$, the
      expectation $\mathbb{E}[v_{T}] = v_{0}\,(\mathbb{E}[r])^{T}$
      grows multiplicatively whenever $\mathbb{E}[r]>1$, but the
      \emph{median} (and any typical sample path) is governed by
      $\mathbb{E}[\log r]$, with median growth controlled by
      $\exp(T\,\mathbb{E}[\log r])$. By Jensen's inequality
      $\mathbb{E}[\log r] \le \log \mathbb{E}[r]$, so
      $\mathbb{E}[r]>1$ does not on its own imply
      $\mathbb{E}[\log r]>0$.
\item \emph{Boundary truncation.} \Cref{eq:value-update} contains a
      $\bar{r}_{t}=1$ fallback for epochs with no winning bids
      ($\mathcal{Y}_{t}=\varnothing$); this puts a point mass at
      unity on the distribution of $r_{t}$ that the clean
      theorem does not capture.
\end{itemize}

\paragraph{Empirical bridge.}
\Cref{tab:rbar-empirical} reports the empirical distribution of
$\bar{r}_{t}$ extracted directly from the EDEM-regime experiments of
\cref{sec:experiments-edem}, computed as
$\bar{r}_{t} = v_{t+T}(h)/v_{t}(h)$ at every epoch boundary across
all seeds. Three rows are reported per scenario: mean, median, and
the share of epochs with $\bar{r}_{t}>1$; an additional row reports
$\mathbb{E}[\log \bar{r}_{t}]$, which is the relevant
condition for sample-path exponential growth. The condition
$\mathbb{E}[\log \bar{r}_{t}] > 0$ holds in every regime examined,
with the no-balancer Run~6 the strongest at $0.041$ per epoch.

\begin{table}[h]
\centering
\small
\caption{Empirical distribution of the per-epoch multiplier
$\bar{r}_{t}$ in the EDEM regimes (\cref{sec:experiments-edem}),
across 1500 epochs (10 seeds, 150 epochs/seed). The
empty-$\mathcal{Y}_{t}$ point mass at $1$ shows up as a low
$\Pr(\bar{r}_{t}>1)$ for the balanced regimes; in those cases the
relevant growth quantity is $\mathbb{E}[\log \bar{r}_{t}]$, whose
positivity certifies sample-path exponential growth.}
\label{tab:rbar-empirical}
\begin{tabular}{l c c c c}
\toprule
\textbf{Regime} & $\mathbb{E}[\bar r_t]$ & $\Pr(\bar r_t>1)$ & $\mathbb{E}[\log \bar r_t]$ & median$\,\bar r_t$ \\
\midrule
Run 6 ($C_b=0,\ \bar\sigma=0.15$)            & 1.0429 & 83.5\% & $+0.0411$ & 1.044 \\
Run 7 ($C_b=+1,\ \bar\sigma=0.15$)           & 1.0104 & 19.7\% & $+0.0098$ & 1.000 \\
Run 7 ($C_b=-1,\ \bar\sigma=0.15$)           & 1.0153 & 25.4\% & $+0.0144$ & 1.000 \\
Run 8 ($C_b=-1,\ \bar\sigma\!\nearrow$)      & 1.0287 & 24.9\% & $+0.0236$ & 1.000 \\
\bottomrule
\end{tabular}
\end{table}

\Cref{eq:ostat-bias} should therefore be read as the
\emph{mechanism}: the same order-statistic logic that yields the
closed form for the seller-side maximum also produces empirical
$\mathbb{E}[\log \bar{r}_{t}] > 0$ for the buyer-min update we
actually simulate. We do not claim that
\cref{eq:ostat-bias} proves \cref{fig:run6-bubble}; we claim that
the closed-form theorem isolates the structural ingredient
(maximum-order-statistic selection) that the empirical
$\bar{r}_{t}$ inherits, and that \cref{tab:rbar-empirical} confirms
the inheritance.

\subsection{DE as a special case}
\label{sec:framework-de-special}

The Dynamic Equilibrium model of \citet{arbuzov2018de} corresponds to
the following restriction of EDEM:
\begin{itemize}[leftmargin=*,nosep]
\item Estimation: $\varepsilon \sim \mathcal{U}[-\sigma_{i}, +\sigma_{i}]$,
      with $\sigma_{i} \sim \mathcal{U}[0, \bar{\sigma}]$ across agents
      (decimal convention; \cref{sec:framework-estimation}).
\item Schedules: $Q_{s}, Q_{d}$ are restored toward linear-in-price
      targets $\hat{Q}_{s}(p) = a_{s} + b_{s} p$,
      $\hat{Q}_{d}(p) = a_{d} + b_{d} p$ once per balance period $T_{B}$.
\item Clearing: $A$ is the rolling average of the last $W$ completed
      sales. Let $k(t)$ count the completed sales up to tick $t$ and
      $P^{\mathrm{sale}}_{j}$ denote the price of the $j$-th sale; then
      \begin{equation}
      \label{eq:price-rolling}
      p_{t} \;=\; \frac{1}{m_{t}} \sum_{j=k(t)-m_{t}+1}^{k(t)} P^{\mathrm{sale}}_{j},
      \qquad m_{t} = \min(W, k(t)).
      \end{equation}
\end{itemize}
This is the formulation that \cref{sec:experiments-de} simulates. The
sale-event indexing of \cref{eq:price-rolling} matters: ticks without
completed sales contribute nothing, and the average is taken over
realised events rather than time. The implementation maintains the
window as a fixed-length deque of the last $W$ sale prices.

The speculative-market EDEM variant of \cref{sec:experiments-edem}
replaces \cref{eq:price-rolling} with a multiplicative per-epoch
update on each home's value. Define:
\begin{align*}
\mathcal{W}_{b}(t) &= \{s \in \mathcal{S}_{t} : b \text{ holds the winning bid on } s \text{ during the epoch ending at } t\}, \\
\mathcal{Y}_{t}    &= \{b : \mathcal{W}_{b}(t) \neq \varnothing\}.
\end{align*}
The per-epoch update is then
\begin{equation}
\label{eq:value-update}
v_{t+T}(h) \;=\; v_{t}(h)\,\cdot\,\bar{r}_{t} \quad \forall h \in G,
\qquad
\bar{r}_{t} \;=\;
\begin{cases}
\dfrac{1}{|\mathcal{Y}_{t}|}\,\displaystyle\sum_{b \in \mathcal{Y}_{t}} \min_{s \in \mathcal{W}_{b}(t)} \dfrac{\beta_{b,s}^{(t)}}{v_{t}(h(s))}
& \text{if } \mathcal{Y}_{t} \neq \varnothing, \\[1.6ex]
1 & \text{if } \mathcal{Y}_{t} = \varnothing.
\end{cases}
\end{equation}

The update is uniform across all homes in $G$. This is a strong
\emph{propagation assumption}: real-estate ``comp'' signals
empirically diffuse non-uniformly, weighted by neighbourhood,
amenities, and time-since-sale. Adopting a uniform multiplier keeps
the model focused on the order-statistic mechanism of
\cref{sec:framework-asymmetry} rather than on diffusion geometry; we
flag it as a candidate target for extension in \cref{sec:ext-clearing}.

\subsection{Walrasian and Miller cases as limits}
\label{sec:framework-walras-miller}

Setting $\sigma_{i} \equiv 0$ for all $i$ collapses
\cref{eq:estimation} to $E_{i,t}(h) = v_{t}(h)$: every agent agrees
on the value of every home. The order-statistic drift of
\cref{eq:ostat-bias} vanishes ($\bar{r}_{t} \equiv 1$) and the
rolling clearing in \cref{eq:price-rolling} delivers a constant
price set by $\hat{Q}_{s}(p^{*}) = \hat{Q}_{d}(p^{*})$. Under the
additional assumption that the matching process and balancer
introduce no frictions (instantaneous adjustment, costless
encounters), EDEM approaches the Walrasian fixed-point equilibrium
as a limiting case. Spatial frictions, patience, and the
finite-population integerization of \cref{eq:Qs-update,eq:Qd-update}
prevent exact equivalence at finite scales.

If we instead retain heterogeneous $\sigma_{i}$ but freeze the
dynamics ($T_{B} \to \infty$, single-period sale, no value update),
we recover a \emph{Miller-style static premium mechanism}: the
$N$ shares are assigned to the $N$ most optimistic of the
$K > N$ potential investors, so the clearing price reflects the
optimism of the marginal optimist rather than the mean valuation.
Exact equivalence to the model in \citet{miller1977} requires
short-selling restrictions and the fixed-supply assumption; the
mechanism here is the same upper-tail selection but in a different
institutional framing.

The two regimes that empirical studies of divergence of opinion have
struggled to reconcile (premium vs.\ discount;
\cref{sec:bg-debate}) appear in EDEM as two ends of a continuum
indexed by the balancer parameters $C_{b}, C_{s}$ and the
ratio of estimation error to balancer strength. \Cref{sec:experiments}
demonstrates this empirically.

\section{Implementation in Mesa}
\label{sec:implementation}

EDEM and DE are implemented in the Python agent-based modelling
framework Mesa~\citep{masad2015mesa}, replacing the earlier NetLogo
prototypes whose NetLogo source is preserved in the
\path{drafts/netlogo/} subdirectory of the same repository. The
choice of Mesa is pragmatic: its turtle--patch primitives map cleanly
onto NetLogo's, but the surrounding Python ecosystem (NumPy, pandas,
matplotlib, pytest) supports batched experimentation, deterministic
seeding, and reproducible figure regeneration that the original
NetLogo code cannot. The Mesa port, the NetLogo source, and the
present paper are released together at
\url{https://github.com/sibmike/dynamic-disequilibrium} --- code under MIT,
paper under CC-BY-4.0.

\subsection{Spatial structure and agents}
\label{sec:implementation-space}

The market is staged on \texttt{mesa.space.MultiGrid(32, 32, torus=True)},
matching the wrapped 32$\times$32 patch world of the original
NetLogo. Per-cell state ($v_{t}(h)$, $v^{*}(h)$, last sale price,
last sale tick) is stored as a lightweight \texttt{Home} dataclass
attached to each grid cell, not as a Mesa Agent --- treating the 1024
cells as Agents would multiply scheduler overhead without benefit,
since cells never schedule behaviour.

Buyers and sellers are Mesa Agents:
\begin{itemize}[leftmargin=*,nosep]
\item \texttt{Seller} is immobile (one per home), holds an ask price
      $a_{t}(s)$ and a patience timer, and accumulates incoming bids
      until the timer elapses.
\item \texttt{Buyer} is mobile; each tick it executes a NetLogo-style
      \emph{wiggle} (uniform heading change in $[-90^{\circ}, +90^{\circ}]$)
      followed by a unit forward step.
\end{itemize}
Bids are NetLogo undirected links in the original; we represent them
as mirrored entries in Python dictionaries on each side (each agent's
\texttt{bids} dict is keyed by the counterparty). The mirror is
maintained in a single helper to avoid drift.

\subsection{Market price and clearing}
\label{sec:implementation-clearing}

For DE, the rolling 25-sale market price of \cref{eq:price-rolling}
is implemented as a \texttt{collections.deque(maxlen=25)} owned by
the model. Sales are committed via a single \texttt{complete\_sale}
method that records the sale, advances the deque, removes both
counterparties from the grid, and triggers the balancer to spawn a
replacement pair (per the NetLogo \texttt{add\_seller} /
\texttt{add\_buyer} idiom).

For the EDEM speculative-market variant, no sales are recorded.
Instead the model maintains a per-tick \texttt{cycle\_counter}
initialised to $\texttt{init\_patience} - 1$. When the counter
reaches zero, the model fires \cref{eq:value-update} on every home in
a single pass and resets the counter. The off-by-one initialisation is
deliberate: it ensures the value update reads buyers' \texttt{is\_yellow}
flags one tick before the buyers' patience timers reset them, matching
the order of operations in the NetLogo \texttt{to go} procedure.

\subsection{Balancer}
\label{sec:implementation-balancer}

DE balances populations every $T_{B}$ ticks by recomputing the linear
target $\hat{Q}_{s}(p), \hat{Q}_{d}(p)$ at the current market price
and adjusting by one agent in each direction. Spawning a replacement
on every sale (the NetLogo behaviour) plus this slow drift toward the
linear target keeps the population near \mbox{$Q_{s}^{*} = Q_{d}^{*}$}
without introducing artificial discrete jumps.

The EDEM balancer implements \cref{eq:Qs-update,eq:Qd-update} in
finite-population form. The fractional balancer coefficient $C_{b}$ is
realised as $\lfloor |C_{b}| \rfloor$ deterministic swaps per epoch
plus one Bernoulli swap with probability $|C_{b}| - \lfloor |C_{b}|
\rfloor$; the sign of $C_{b}$ controls direction. A population floor
prevents the balancer from killing the last agent on either side; the
1-vs-many state that emerges with strong $|C_{b}|$ is kept as a feature
of the model rather than papered over (\cref{sec:exp-balancer-sweep}).

\subsection{Bid-acceptance rule (Cond.~2)}
\label{sec:implementation-cond2}

The original \citet{arbuzov2018de} prose stated the buyer's
bid-acceptance rule with the strict inequality
$\beta < \mathrm{mean}(\mathcal{B}_{b})$, while the accompanying
NetLogo source compared with the opposite sign. Empirically the
prose-as-written rule causes a runaway price collapse, since it
selects every sale to clear at the buyer's lowest bid; only the
NetLogo-as-coded rule reproduces the stable equilibrium reported in
the original Run~1. We adopt \cref{eq:cond2} (the NetLogo-as-coded
rule, restated economically) as the canonical Cond.~2 and document
this discrepancy explicitly in the source. The Mesa implementation
exposes both rules as a parameter (\texttt{accept\_rule="netlogo"}
or \texttt{"prose"}) for sensitivity analysis.

\subsection{Determinism, parallelism, and reproducibility}
\label{sec:implementation-repro}

The Mesa \texttt{Model.rng} is seeded explicitly per run, and all
agent-side randomness (epsilon draws, heading wiggles, balancer
Bernoulli) draws from this single stream. Each experiment script in
\path{python_simulation/experiments/} runs $N \ge 8$ independent seeds
of the same parameter vector and stacks the per-tick model reporters
into a single Parquet dataset; figures plot the median across seeds
plus the 10th--90th percentile band. Pytest unit tests
(\path{python_simulation/tests/}) cover the rolling market price, both
Cond.~2 rules, the linear and fractional balancers, the EDEM
epoch-timing trick, and the $C_{b}$-sign inversion.

\subsection{Per-tick pseudocode}
\label{sec:implementation-pseudocode}

The complete tick is:

\begin{lstlisting}[language=Python]
def step(self):  # Model.step
    self.agents.shuffle_do("step")     # buyers + sellers, randomised
    self.balancer.step(self)           # DE: every T_B; EDEM: per-epoch
    self.datacollector.collect(self)
\end{lstlisting}

\noindent
where \texttt{Buyer.step} runs \emph{buy + wiggle + move} in NetLogo
order and \texttt{Seller.step} runs the patience-timer logic
(timer decrement; on timeout, either drop the highest bid below ask
and lower the ask, or offer to sign the highest bidder, applying
\cref{eq:cond2}). The full source is under 600 lines including
docstrings.

\section{Experiments and Results}
\label{sec:experiments}

We report eight primary controlled experiments grouped into two
regimes, plus a sensitivity-grid sweep (Run~9,
\cref{sec:exp-sensitivity}) that anchors the conclusions across a
30-cell parameter grid.
\Cref{sec:experiments-de} (Runs~1--5) uses the Dynamic Equilibrium
specialisation of EDEM with linear supply and demand schedules,
discrete sales, and a rolling-mean clearing price; these isolate the
effects of estimation noise, patience, agent density, and shocks
under a textbook population balancer. \Cref{sec:experiments-edem}
(Runs~6--8) uses the speculative-market EDEM variant with
multiplicative value updates; these illustrate the bubble, balancer
sign, and rising-error regimes that the static
divergence-of-opinion literature cannot reach.

Unless stated otherwise, all runs use a $32 \times 32$ toroidal grid,
balance period $T_{B} = 100$ ticks, sale window $W = 25$ sales, and
the bid-acceptance rule \cref{eq:cond2}. Each experiment is replicated
across at least eight independent seeds; figures plot the median across
seeds plus the 10th--90th percentile band, with individual seed
trajectories overlaid faintly. Full parameter tables are in
\cref{app:params}.

\subsection{Dynamic Equilibrium experiments}
\label{sec:experiments-de}

\subsubsection{Run 1 -- Stable equilibrium under favourable conditions}
\label{sec:exp-stable}

\begin{figure}[t]
\centering
\includegraphics[width=\linewidth]{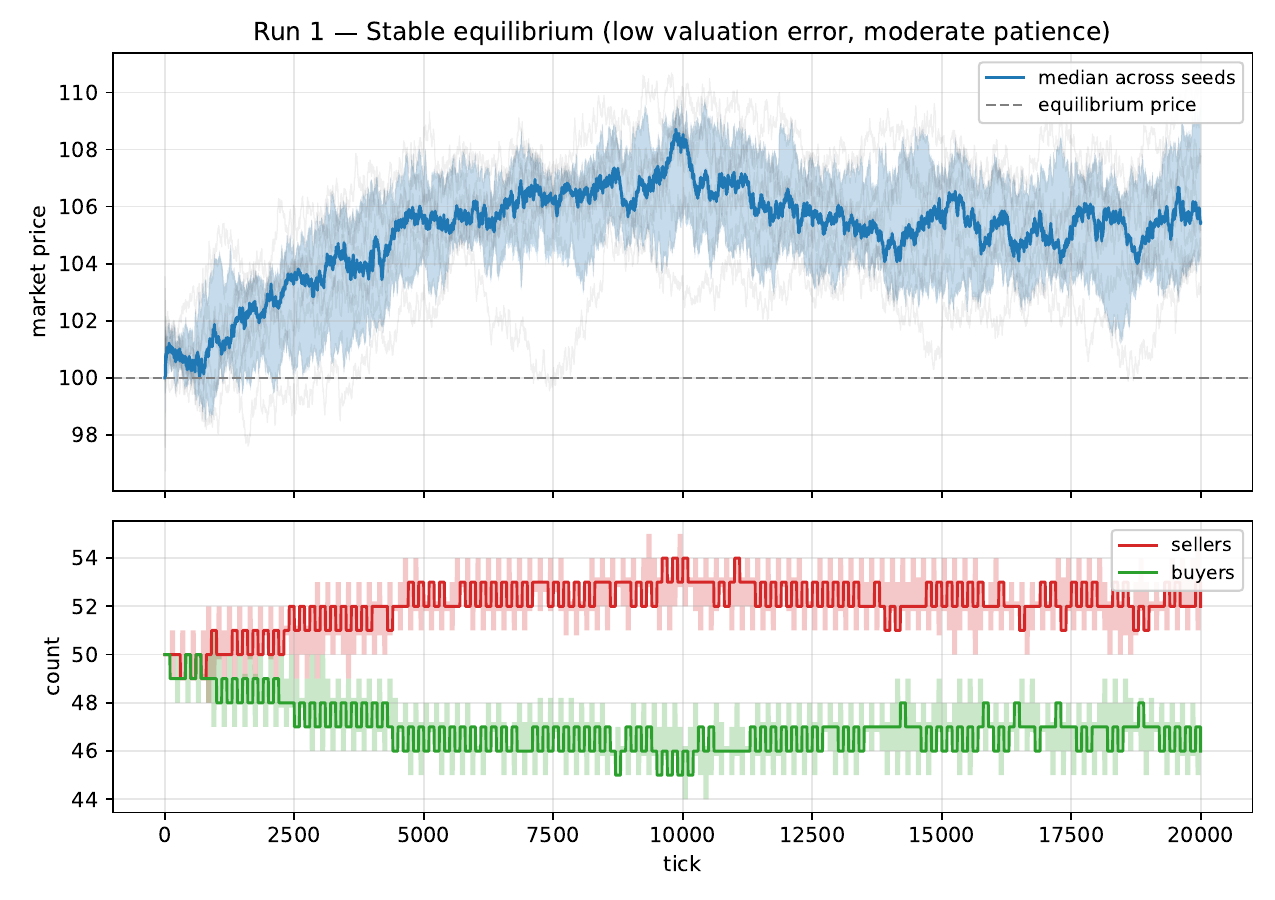}
\caption{Run~1 -- Stable equilibrium with low valuation error and
moderate patience. Top: market price (median, with 10--90 percentile
band; faint individual-seed traces). Bottom: agent counts. Dashed
line: textbook equilibrium price.}
\label{fig:run1-stable}
\end{figure}

We begin from the textbook equilibrium $p^{*} = 100$, $q^{*} = 50$
defined by $\hat{Q}_{s}(p) = 0.5\,p$ and
$\hat{Q}_{d}(p) = 100 - 0.5\,p$, with maximum estimation error
$\bar{\sigma} = 5$ percent, patience drawn uniformly on $[0, 50]$,
and balance period $T_{B} = 100$. \Cref{fig:run1-stable} shows the
median market price hovering close to $p^{*}$: across ten seeds and
20{,}000 ticks, the maximum upward deviation is $+10.7\%$ and the
maximum downward deviation $-2.4\%$. The companion buyer/seller
counts oscillate within $\pm 5$ of the equilibrium quantity. The
slight upward bias of the median trajectory is consistent with
\citet{arbuzov2018de}'s reported $-4.9\%/+6.2\%$ envelope at
100{,}000 ticks; we attribute it to the bid-selection asymmetry
discussed in \cref{sec:framework-de-special}, which the rolling
average partially absorbs.

\subsubsection{Run 2 -- Imprecise valuations cause business cycles}
\label{sec:exp-cycles}

\begin{figure}[t]
\centering
\includegraphics[width=\linewidth]{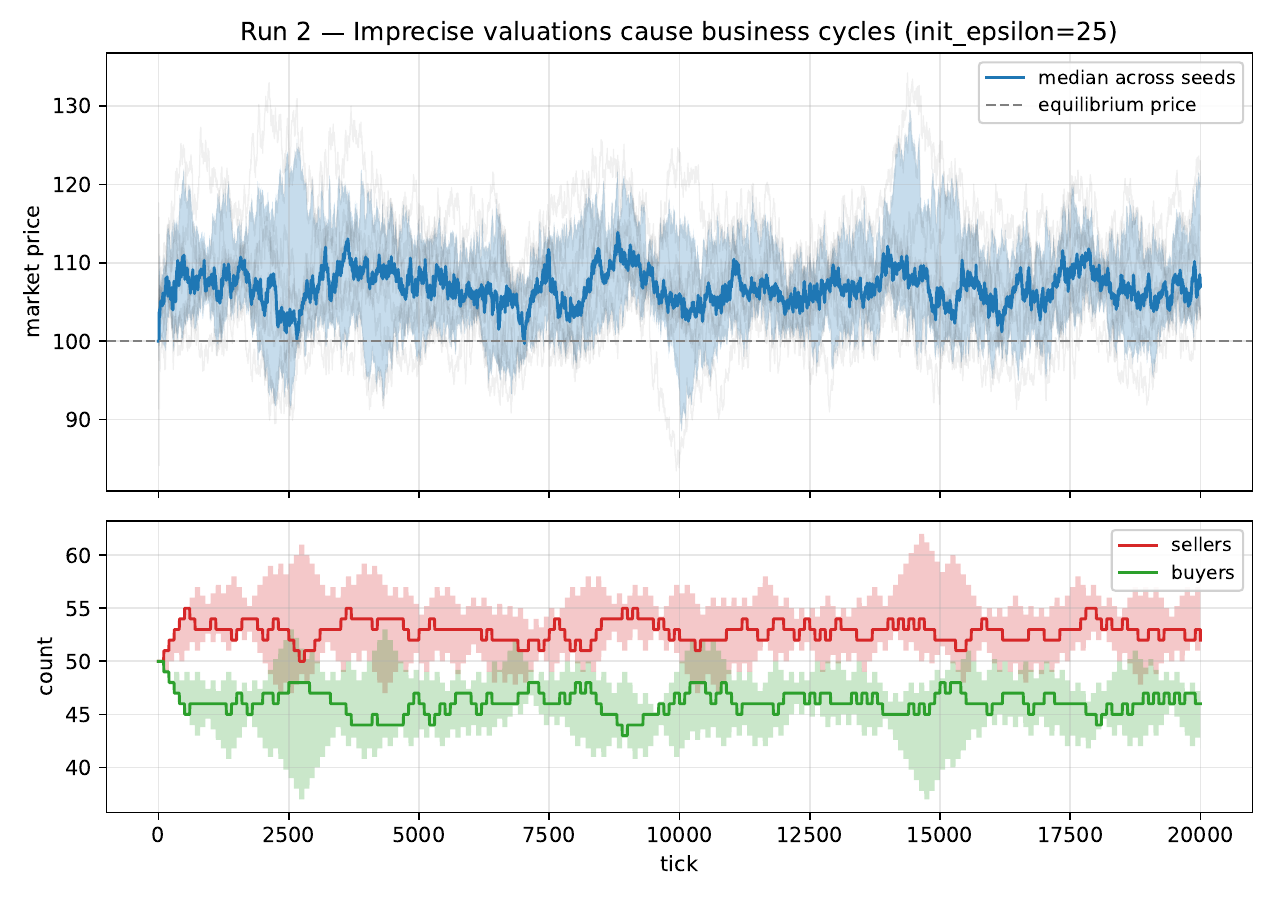}
\caption{Run~2 -- Imprecise valuations (\(\bar{\sigma} = 25\%\))
generate endogenous business cycles. Same conventions as
\cref{fig:run1-stable}.}
\label{fig:run2-cycles}
\end{figure}

Holding all other parameters at Run~1 values and increasing
$\bar{\sigma}$ from $5$ to $25$ percent, the market loses its
narrow-band stability. \Cref{fig:run2-cycles} shows recurrent
oscillations of the market price between roughly $+34\%$ and
$-17\%$ of equilibrium, accompanied by counter-phased oscillations
of buyer and seller populations: episodes of overbidding draw
sellers into the market while buyers retreat, after which the
seller surplus pushes prices back down and the cycle reverses.
This is a pure consequence of estimation noise --- there are no
exogenous shocks --- and it suggests that observed business cycles
in real markets need not require any external driver beyond
heterogeneous misvaluation, provided the divergence of opinion is
large relative to the balancer's adjustment speed.

\subsubsection{Run 3 -- Patience pushes price above equilibrium}
\label{sec:exp-patience}

\begin{figure}[t]
\centering
\includegraphics[width=\linewidth]{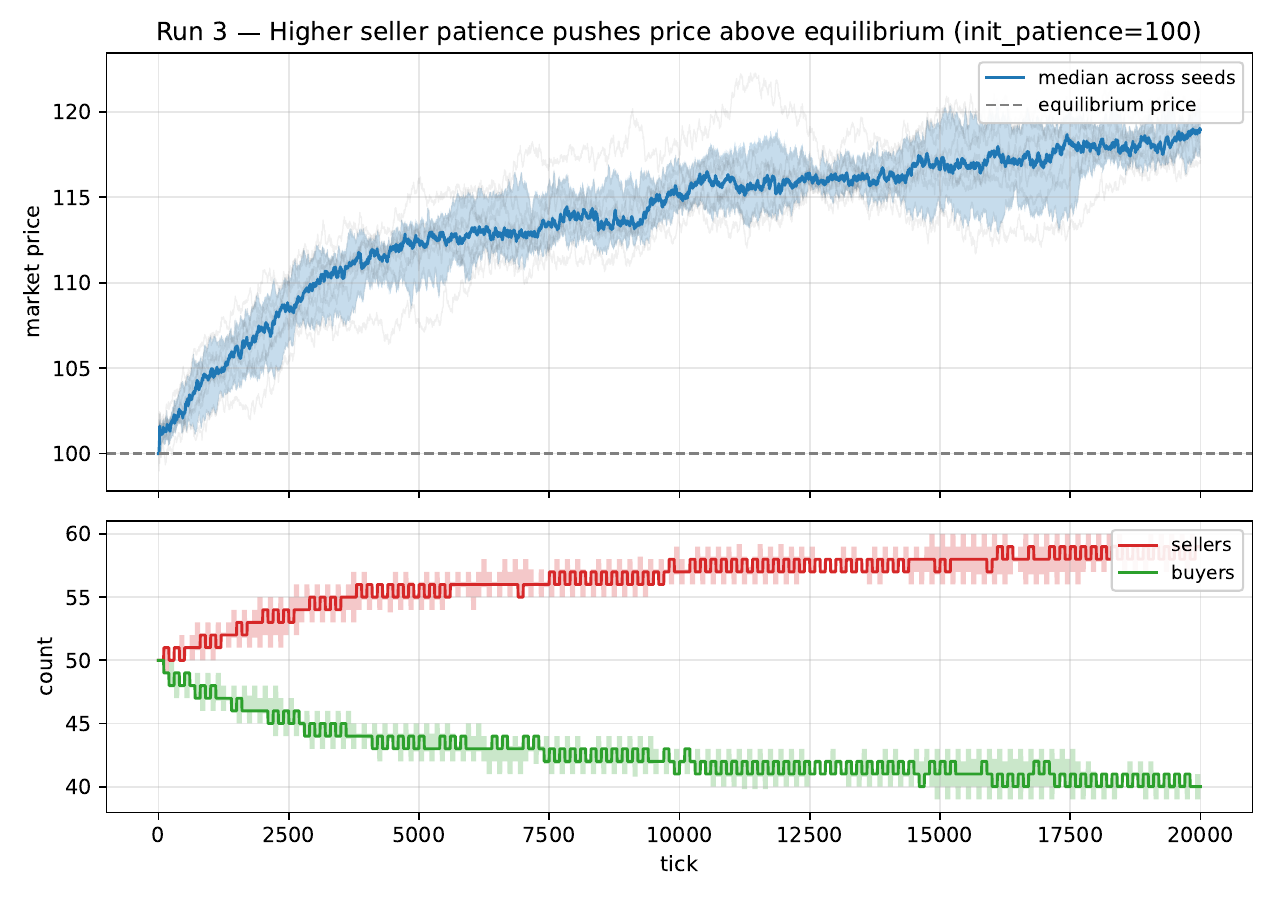}
\caption{Run~3 -- Doubling seller patience to $100$ ticks pushes the
market price about $14\%$ above the textbook equilibrium and
sustains it there. Same conventions as \cref{fig:run1-stable}.}
\label{fig:run3-patience}
\end{figure}

Doubling the maximum seller patience from $50$ to $100$ ticks
(holding $\bar{\sigma}$ at the Run~1 value of $5\%$) produces a
durable upward shift in the market price.
\Cref{fig:run3-patience} shows the median price climbing
monotonically to roughly $+14\%$ above $p^{*}$ over 20{,}000 ticks
and showing no sign of reverting. The seller population settles
at $\sim 57$ (about $+14\%$ of $q^{*}$) and the buyer population
at $\sim 41$ ($-18\%$). Patience increases sellers' time-on-market
and thus their willingness to wait for richer bids; the realised
sale prices tilt up; the rolling average follows.
\citet{arbuzov2018de} reported a similar
$+10\%/+10\%/-20\%$ shift, attributing it to seller forbearance
on stagnating real-estate markets where neither side can quickly
exit.

\subsubsection{Run 4 -- Low agent density pushes price below equilibrium}
\label{sec:exp-density}

\begin{figure}[t]
\centering
\includegraphics[width=\linewidth]{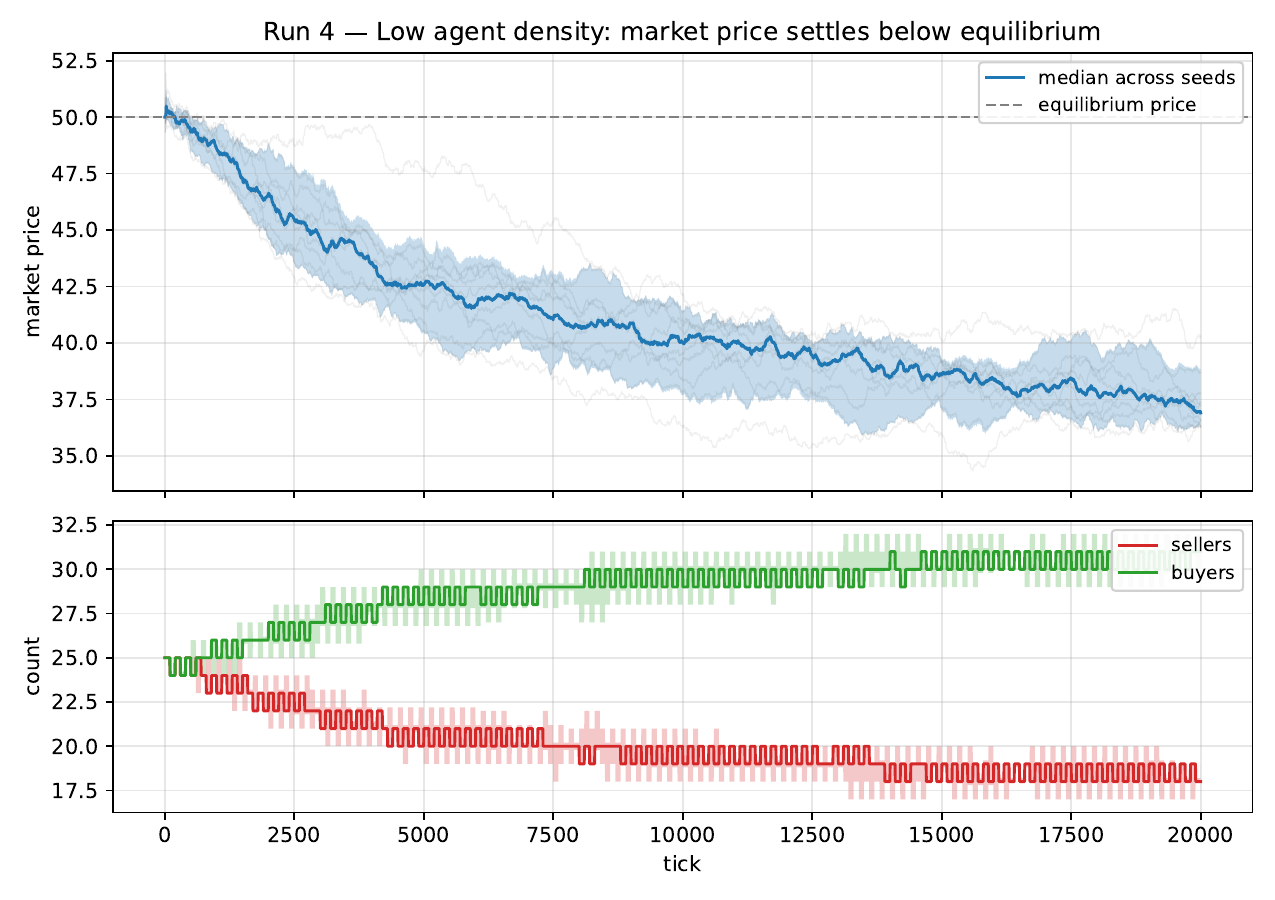}
\caption{Run~4 -- Halving the demand intercept reduces equilibrium
quantity to $25$. Sparse encounters cause sellers to lower ask prices
repeatedly; the realised market price settles about $18\%$ below the
textbook equilibrium. Same conventions as \cref{fig:run1-stable}.}
\label{fig:run4-density}
\end{figure}

Shifting the demand intercept downward by $50$ units yields a new
textbook equilibrium $(p^{*}, q^{*}) = (50, 25)$. With only $\sim
25$ buyers and $\sim 25$ sellers on a $32 \times 32$ grid, encounters
are sparse: a typical buyer's wiggle-and-step trajectory rarely
crosses a seller's home before the seller's patience timer expires.
Sellers therefore lower ask prices repeatedly without receiving
adequate bids, and the realised sale prices --- and hence the rolling
market price --- settle \emph{below} the textbook equilibrium.
\Cref{fig:run4-density} shows a steady downward drift; the
mean realised price across the second half of the simulation is
$p \approx 41$, an $18\%$ shortfall against the textbook
equilibrium of 50, with episodes reaching as low as $p \approx 37$
(a $26\%$ shortfall at the trough). The seller population shrinks
to $\sim 19$ and the buyer population grows to $\sim 30$ as the
balancer chases the depressed price. Patience and density are partial
substitutes (\cref{sec:bg-dynamics}): both proxy for time-on-market.

\subsubsection{Run 5 -- Shock handling and transitional markets}
\label{sec:exp-shock}

\begin{figure}[t]
\centering
\includegraphics[width=\linewidth]{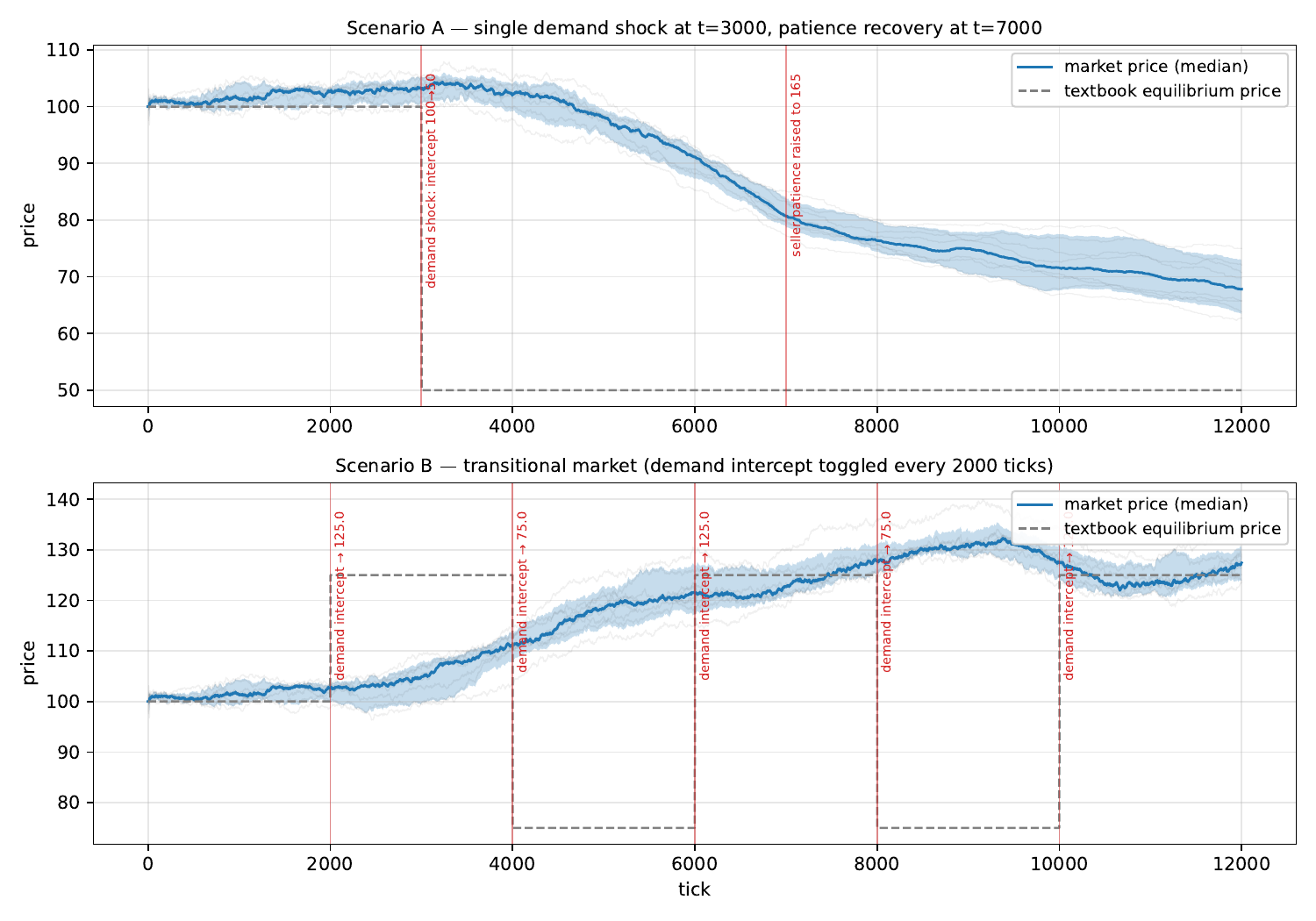}
\caption{Run~5 -- Two shock scenarios. Top: a single demand shock at
$t = 3000$ followed by a patience boost at $t = 7000$; the market
absorbs the shock slowly and has not fully adjusted by the end of
the simulation. Bottom: a transitional market in which the demand
intercept toggles every $2000$ ticks. The market price never
converges to either textbook equilibrium. Red dashed verticals mark
shock events.}
\label{fig:run5-shock}
\end{figure}

The previous four runs evolve under fixed parameters. Run~5
introduces shocks via the
\texttt{set\_demand}/\texttt{set\_supply} hooks
(\cref{sec:implementation-pseudocode}) and demonstrates how the model
absorbs them. \Cref{fig:run5-shock} reports two scenarios.

\paragraph{Scenario A (single shock).} The market begins in the Run~1
equilibrium. At $t = 3000$ the demand intercept drops to $50$,
collapsing the textbook equilibrium to $p^{*} = 50$. The realised
price descends slowly --- it has not finished adjusting by $t = 7000$
when seller patience is raised to $165$. The combination of a
lowered demand and elevated patience continues to drive the market
downward (because the price is still adjusting from the previous
$p^{*} = 100$), and even by $t = 12000$ the realised price has not
settled. Doc 2's claim that high patience would lift the price
\emph{back to} $p^{*} = 50$ presumes a fully-adjusted starting state
that is not reached within twelve thousand ticks; the figure makes
visible the more general \citet{arbuzov2018de} observation that
``adjustment to single shock takes a long time.''

\paragraph{Scenario B (transitional market).} The demand intercept
toggles between $125$ and $75$ every $2000$ ticks. The textbook
equilibrium accordingly steps between $p^{*} = 125$ and $p^{*} = 75$,
but the realised market price never settles at either: each shock
is absorbed only partially before the next arrives. The figure
confirms a market \emph{in constant transition from one unknown
state to another} --- the empirical foothold for
\cref{sec:framework-schedules}'s claim that supply and demand are
sample paths of a stochastic process, not deterministic schedules.

\subsection{Speculative-market EDEM experiments}
\label{sec:experiments-edem}

The remaining three runs use the EDEM variant of
\cref{sec:framework-de-special} with multiplicative value updates
\cref{eq:value-update}. We track the dimensionless ratio
$v_{t}(h) / v^{*}(h)$ averaged over all homes; a value of $1$
indicates fair valuation, values above $1$ indicate the market has
priced homes above their unobservable fair value.

\subsubsection{Run 6 -- Speculative bubble at $C_{b} = 0$}
\label{sec:exp-bubble}

\begin{figure}[t]
\centering
\includegraphics[width=\linewidth]{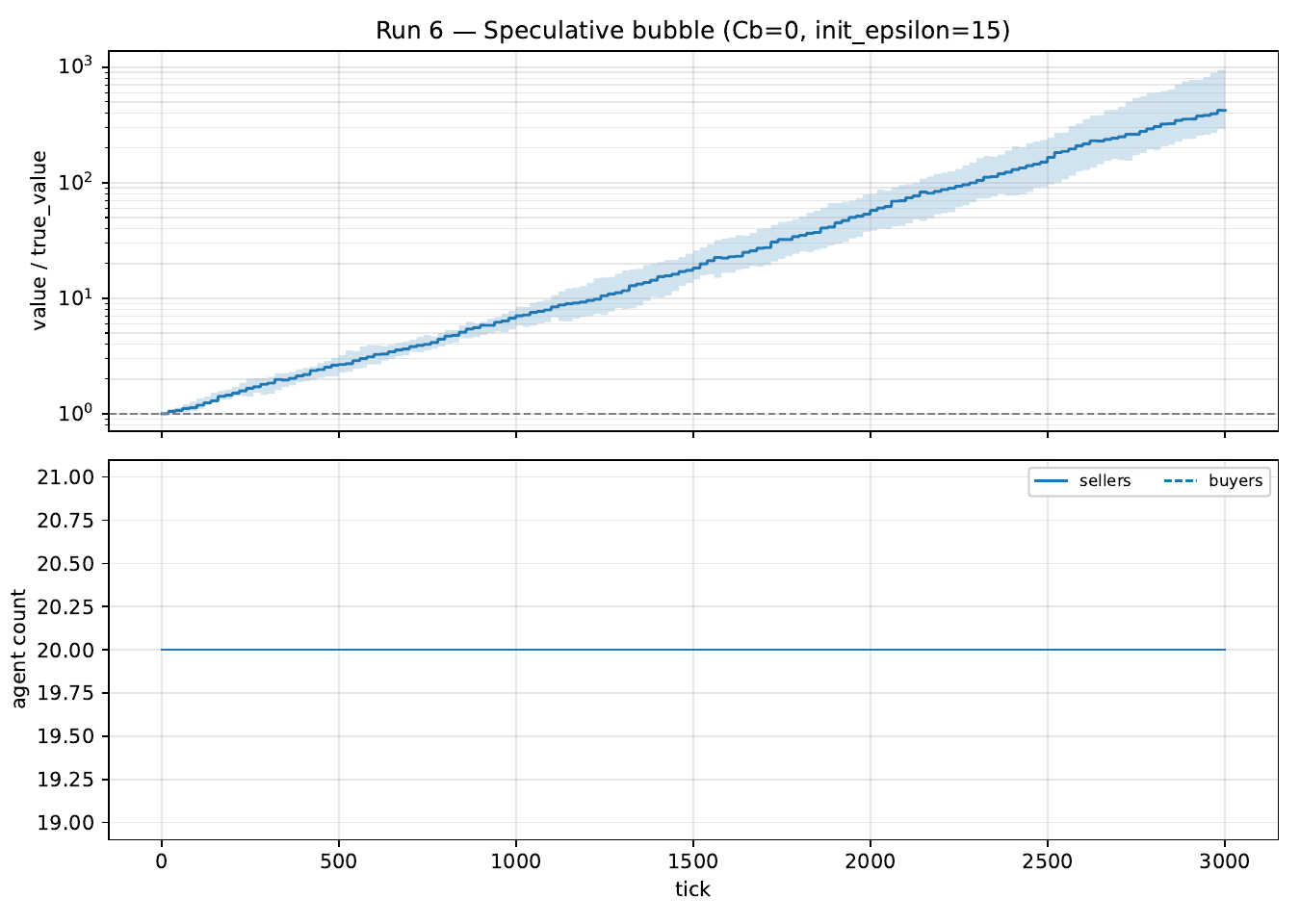}
\caption{Run~6 -- Speculative bubble at $C_{b} = 0$,
$\bar{\sigma} = 15\%$. Top: dimensionless market value $v_{t}/v^{*}$
on a logarithmic axis; the trajectory is approximately linear in
$\log v$, indicating exponential growth. Bottom: agent populations
remain at $(20, 20)$.}
\label{fig:run6-bubble}
\end{figure}

With balancer disabled ($C_{b} = 0$) and divergence of opinion
$\bar{\sigma} = 15\%$, the market exhibits a clean log-linear bubble.
\Cref{fig:run6-bubble} shows the median value-to-true ratio
multiplying by $\sim 421\times$ over $3000$ ticks ($150$ epochs),
with the band tightly enclosing the median: the bubble is
deterministic in shape and stochastic only in slope. Buyer and
seller populations are frozen at $20$ each, since $C_{b} = 0$
admits no balancer adjustment. The mechanism is the bid-selection
asymmetry of \cref{sec:framework-de-special}: each epoch's average
winning ratio $\bar{r}_{t}$ is $> 1$ in expectation because winning
bids are the maxima of multiple noisy draws, and \cref{eq:value-update}
compounds this multiplicatively over epochs.

\subsubsection{Run 7 -- Balancer sign shapes (but cannot eliminate)
                       the bubble}
\label{sec:exp-balancer-sweep}

\begin{figure}[t]
\centering
\includegraphics[width=\linewidth]{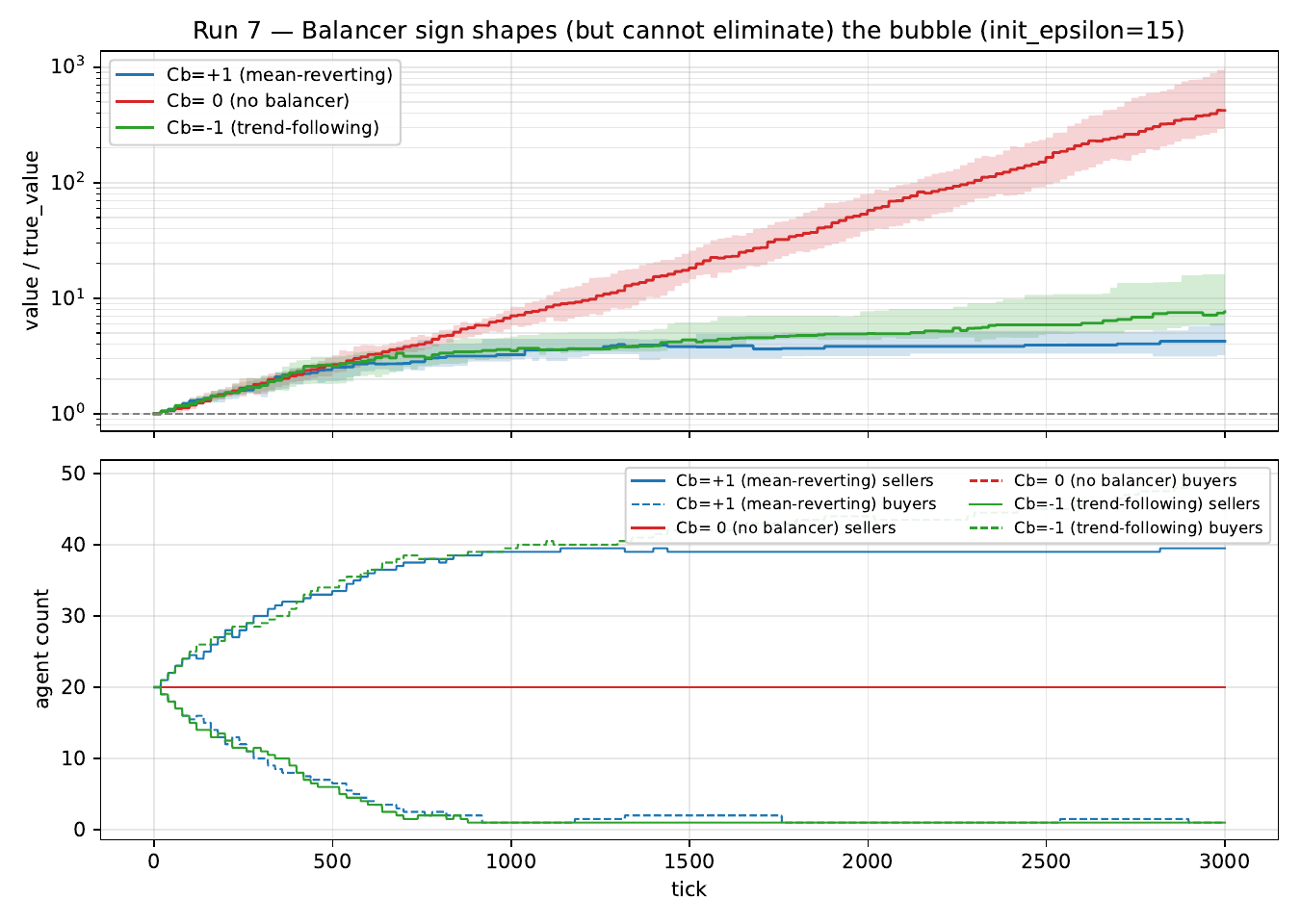}
\caption{Run~7 -- Balancer-coefficient sweep at $\bar{\sigma} = 15\%$.
Three regimes are overlaid: mean-reverting ($C_{b} = +1$), no
balancer ($C_{b} = 0$), and trend-following ($C_{b} = -1$). The
no-balancer case dominates; the trend-following case beats the
mean-reverting case but is itself bounded above by the no-balancer
case. Bottom panel shows the population swings induced by each
balancer.}
\label{fig:run7-balancer-sweep}
\end{figure}

Holding $\bar{\sigma}$ at $15\%$ and varying $C_{b} \in \{+1, 0, -1\}$
yields three qualitatively distinct trajectories overlaid in
\cref{fig:run7-balancer-sweep}. The mean-reverting balancer
($C_{b} = +1$) dampens the bubble most aggressively: the value ratio
plateaus near $4 \times$ as the seller population grows to roughly
$40$ and the buyer population shrinks to one. The trend-following
balancer ($C_{b} = -1$) admits a larger bubble than $+1$, ending
near $7.7\times$, with mirrored populations (one seller, $\sim 40$
buyers). The no-balancer case ($C_{b} = 0$) dominates both, ending
at $\sim 421\times$.

This last observation refines the framing in
\citet{arbuzov2018edem}, which described the trend-following case
as the most explosive (the ``Silicon Valley real estate'' or
``bitcoin'' regime). In the Mesa simulation the most explosive case
is the no-balancer one: removing one population pole entirely (as
trend-following does) reduces the multiplicative pressure on each
remaining seller, because there are simply fewer sellers whose
values can be inflated. The trend-following case is more aggressive
than mean-reverting, but neither is as aggressive as no balancer.

The companion finding is that no setting of $C_{b}$ in this
parameter grid restores $v_{t} / v^{*} = 1$. Bid-selection
asymmetry introduces a positive bias that even a strong
mean-reverting balancer cannot fully cancel within the timescales
considered.

\subsubsection{Run 8 -- Trend-following balancer with rising error}
\label{sec:exp-double-exp}

\begin{figure}[t]
\centering
\includegraphics[width=\linewidth]{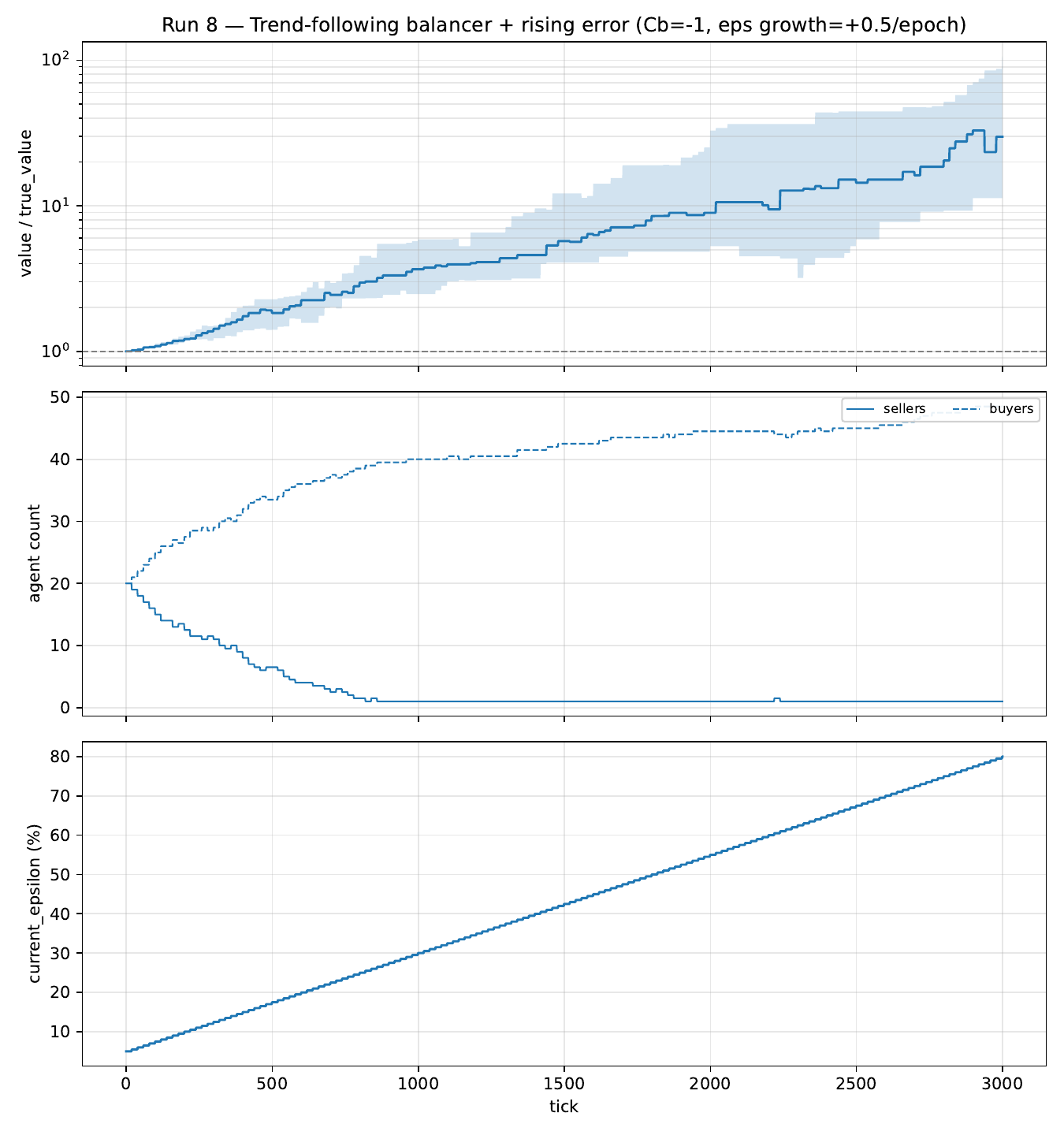}
\caption{Run~8 -- Trend-following balancer ($C_{b} = -1$) combined
with linearly-growing divergence of opinion
($\bar{\sigma}: 5 \to 80$). Top panel: market value on a log scale.
Middle: agent populations. Bottom: $\bar{\sigma}$ over time. The
trajectory is sustained exponential rather than visibly
double-exponential.}
\label{fig:run8-double-exp}
\end{figure}

Doc 2's stylised ``bitcoin'' or ``Silicon Valley'' regime combines
trend-following ($C_{b} = -1$) with a divergence of opinion that
itself grows over time, modelling the historical observation that
late-stage bull markets see widening dispersion in analyst forecasts.
We initialise $\bar{\sigma}_{0} = 0.05$ (i.e.\ $5\%$) and let
$\bar{\sigma}_{t+T} = \bar{\sigma}_{t} + 0.005$ each epoch
(equivalently, $+0.5$ percentage points), so $\bar{\sigma}$ reaches
$0.80$ ($80\%$) by the end of the run.
\Cref{fig:run8-double-exp} shows the resulting trajectory: the
value ratio climbs to $\sim 30 \times$ over $3000$ ticks. The
trajectory is super-linear in linear time but visibly concave-down
in log-time, indicating sustained exponential rather than
double-exponential growth at this parameter setting; we conjecture
that genuine double-exponential acceleration would emerge for
larger $\bar{\sigma}$ growth rates than we explore here. The
population panel mirrors Run~7's $C_{b} = -1$ behaviour: sellers
drained to one, buyers saturated at $\sim 44$.

\subsubsection{Run 9 -- Sensitivity grid}
\label{sec:exp-sensitivity}

\begin{figure}[t]
\centering
\includegraphics[width=\linewidth]{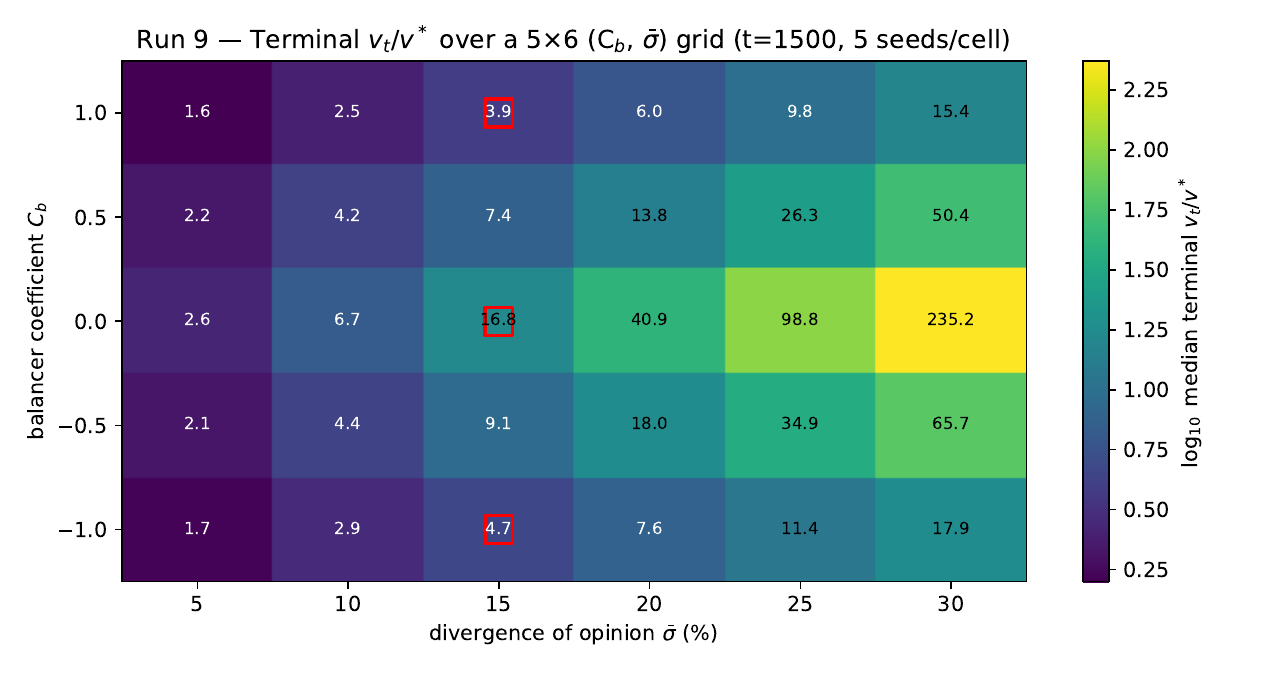}
\caption{Run~9 -- Sensitivity grid: median terminal $v_{t}/v^{*}$
over a $5 \times 6$ sweep of balancer coefficient $C_{b}$ versus
divergence of opinion $\bar{\sigma}$, with 5 seeds per cell at
$t = 1500$. Colour: $\log_{10}$ of the cell median; cell text:
linear value. Red squares mark the cells corresponding to Runs~6
and~7 ($C_{b} \in \{-1, 0, +1\}$ at $\bar{\sigma} = 15\%$). The
order-statistic drift produces $v_{t}/v^{*} > 1$ in every cell.}
\label{fig:run9-sensitivity}
\end{figure}

The eight primary runs reported above each fix a single parameter vector. To
rule out the possibility that the order-statistic drift of
\cref{sec:framework-asymmetry} is an artefact of one well-chosen
calibration, we sweep $C_{b}$ and $\bar{\sigma}$ over a $5 \times 6$
grid (\(C_{b} \in \{-1, -0.5, 0, +0.5, +1\}\),
$\bar{\sigma} \in \{5, 10, 15, 20, 25, 30\}\%$), running five seeds per
cell for $1500$ ticks. \Cref{fig:run9-sensitivity} reports the
median terminal $v_{t}/v^{*}$ across seeds, on a $\log_{10}$ colour
scale.

Three observations stand out:

\begin{enumerate}[leftmargin=*,nosep]
\item \textbf{The drift is universal in this range.} Every cell of the
grid has $v_{t}/v^{*} > 1$, and 100\% of cells exceed
$1.5 \times$. The order-statistic mechanism is not a corner-case.
\item \textbf{Bubble territory is large.} 43\% of cells exceed
$10 \times$ within $1500$ ticks; growth at $\bar{\sigma} = 30\%$
reaches $235 \times$ in the no-balancer column.
\item \textbf{$C_{b} = 0$ dominates within each $\bar{\sigma}$ column.}
This confirms the Run~7 finding (\cref{sec:exp-balancer-sweep}) at
five distinct $\bar{\sigma}$ values: removing the balancer is more
explosive than either mean-reverting or trend-following balancers,
across the whole grid.
\end{enumerate}

The grid does not constitute a closed-form proof that the drift
survives \emph{all} parameter regimes; in particular, very large
$|C_{b}|$ values that drain one side of the population to a single
agent within a few epochs may behave differently. Section
\ref{sec:disc-balancer} discusses the limits of the linear-balancer
intervention more carefully.

\subsection{Summary across runs}
\label{sec:exp-summary}

\Cref{tab:experiment-summary} consolidates the end-of-run statistics.
The eight primary runs sweep three parameter axes: estimation noise
($\bar{\sigma}$), seller patience, and agent density (Runs 1--4); a
shock schedule (Run~5); and the EDEM balancer parameter $C_{b}$ with
a fixed or growing $\bar{\sigma}$ (Runs 6--8). Together they delineate
six qualitatively distinct steady-state regimes -- band-stable,
business-cycle, persistent overshoot, persistent undershoot, runaway
exponential, and constant transition -- all reachable from the same
underlying agent ruleset.

\begin{table}[t]
\centering
\small
\caption{Eight EDEM runs: parameters and observed outcomes
at the end of each simulation. Median across $\ge 8$ seeds.
DE runs report median deviation of the rolling market price from the
textbook equilibrium; EDEM runs report the dimensionless
$v_{t}/v^{*}$ ratio.}
\label{tab:experiment-summary}
\begin{tabular}{l l l l}
\toprule
\textbf{Run} & \textbf{Manipulated parameter} & \textbf{Outcome metric} & \textbf{Regime} \\
\midrule
1 & $\bar{\sigma} = 5\%$, patience $\le 50$       & $-2\%$ to $+11\%$ band       & band-stable        \\
2 & $\bar{\sigma} = 25\%$                         & $-17\%$ to $+34\%$, periodic & business-cycle     \\
3 & patience $\le 100$                            & $+14\%$ persistent           & overshoot          \\
4 & demand intercept $-50$                        & $-18\%$ persistent           & undershoot         \\
5 & shock at $t=3000$, then re-shocks             & price chases moving target   & transitional       \\
\midrule
6 & EDEM, $C_{b} = 0$, $\bar{\sigma}=15\%$        & $v_{t}/v^{*} = 421\times$    & runaway bubble     \\
7 & EDEM, $C_{b} \in \{-1, 0, +1\}$               & $7.7,\,421,\,4.2\times$ (in $C_b$ order)    & balancer-shaped    \\
8 & EDEM, $C_{b} = -1$, $\bar{\sigma}: 5\to80$    & $v_{t}/v^{*} = 30\times$     & sustained super-linear \\
\bottomrule
\end{tabular}
\end{table}

\section{Discussion}
\label{sec:discussion}

The eight primary experiments plus the Run-9 sensitivity grid
together support three substantive claims about
how real markets generate the price patterns that empirical research
has found contradictory.

\subsection{Time-on-market is the missing third dimension}
\label{sec:disc-tom}

Runs~3 and~4 show that two distinct manipulations --- raising seller
patience and lowering agent density --- both produce stable
deviations from the textbook equilibrium price. Their effect signs
are opposite (Run~3 overshoots, Run~4 undershoots), but their
mechanism is the same: both alter the typical \emph{time on market}
of an unsold home. Patience increases the upper bound on time on
market directly; density increases it indirectly by decreasing
encounter rates per unit time. In each case the mismatch between the
balancer's price-to-quantity correction and the underlying microstructural
adjustment shows up as a persistent equilibrium offset.

This suggests that the textbook supply-and-demand pair $\hat{Q}_{s}(p)$
and $\hat{Q}_{d}(p)$ omits a state variable. A more accurate
characterisation would be $\hat{Q}_{s}(p, \tau)$ and
$\hat{Q}_{d}(p, \tau)$ where $\tau$ is expected time on market.
Empirically, real-estate datasets that include days-on-market
information vastly outperform those that do not for short-run
forecasting; our model offers a generative explanation for that
finding.

\subsection{Bid-selection asymmetry, not estimation bias, drives bubbles}
\label{sec:disc-bias}

The estimation function \cref{eq:estimation} has zero mean error: an
agent's expected valuation is exactly the home's current value. Yet
the EDEM speculative regime (Runs 6--8) produces unbounded upward
drift even with this unbiased estimator. The mechanism is not
behavioural bias but a statistical asymmetry: each seller selects the
\emph{maximum} bid received, and the maximum of $N$ noisy estimates
of a fair value is, in expectation, larger than the fair value
itself. \cref{eq:value-update} compounds this asymmetry
multiplicatively across epochs.

This refines the standard ``divergence of opinion'' story.
\citet{miller1977}'s static premium emerges from constrained supply:
the $N$ shares are sold to the $N$ most optimistic of $K > N$
investors. EDEM's dynamic premium emerges from \emph{repeated}
constrained supply: each epoch's winners feed back into the next
epoch's anchor value. The static premium is a corner of the dynamic
premium, recovered when the feedback loop is severed.

A practical implication is that empirical premium-vs-discount
findings depend strongly on the sample's position along this
feedback trajectory, which short-window studies cannot identify.
Two studies that find opposite signs may both be correct
characterisations of distinct epochs in the same underlying process.

\subsection{Mean-reverting balancers cannot fully cancel the asymmetry}
\label{sec:disc-balancer}

Run~7's three-way comparison shows that no setting of the EDEM
balancer coefficient $C_{b}$ within the parameter grid we examined
restores $v_{t}/v^{*} = 1$, and the 30-cell sweep of Run~9
(\cref{sec:exp-sensitivity}) reproduces this finding at every
combination of $C_{b}$ and $\bar{\sigma}$ tested. The
mean-reverting case ($C_{b} = +1$) plateaus near $4 \times$ at
$\bar{\sigma} = 0.15$; the trend-following case ($C_{b} = -1$)
continues drifting upward; the no-balancer case dominates both. A
corollary, qualified by the parameter range surveyed, is that
\emph{linear policy interventions cast as balancer adjustments will
not, on the evidence assembled here, eliminate persistent
mispricing.} We do not claim this as an impossibility result --- the
parameter grid is finite, and we have not explored extreme
$|C_{b}|$ values where the population dynamics qualitatively change.
The asymmetry is structural to the bidding mechanism; to neutralise
it, an intervention would have to change either the selection rule
(e.g., uniform random matching instead of max-bid) or the per-epoch
update rule (e.g., a winsorised mean instead of the arithmetic
mean). \Cref{sec:extensions} sketches both alternatives.

The companion observation --- that the no-balancer case dominates
the trend-following case in our simulation --- refines the framing
in \citet{arbuzov2018edem}, which described the trend-following
``Silicon Valley real estate'' regime as the most explosive. That
framing turns out to overstate the role of the balancer's sign
relative to its presence. We attribute the discrepancy to the
agent-pool depletion that strong $|C_{b}|$ induces: trend-following
removes sellers, but the remaining sellers are all that the
multiplicative update rule can act on.

\subsection{Implications for valuation algorithms}
\label{sec:disc-ml}

The same mechanism that drives EDEM's exponential bubble in Run~6
operates in any pricing system whose inputs are anchored to recent
market-clearing prices. A machine-learning real-estate valuation
algorithm trained on historical sale prices estimates not the home's
fair value but the typical winning bid for a similar home --- which
is, by construction, the maximum of the relevant noisy estimates,
and therefore systematically above the fair value. Deploying such an
algorithm at scale risks providing the market with a coordinating
signal that pulls realised winning bids upward, which then feeds
back into the next training cycle.

We do not advance EDEM as an explanation for any specific real-world
mispricing event. The 2021 wind-down of Zillow's algorithmic-iBuying
programme \citep{zillow2021ibuying} is one episode in which a
historical-sale-price-anchored valuation system at scale incurred
substantial losses; the public record is consistent with EDEM's
order-statistic prediction but is also consistent with several other
explanations (timing of housing-market regime changes, operational
mispricing, capital-allocation constraints), and we have not
attempted to discriminate between them. The robust theoretical
finding is more modest: an unbiased ML valuation trained on
historical clearing prices is, by construction, fitting the
upper-order statistic rather than the population mean of plausible
sale prices, and at deployment time it functions as a
positively-biased estimator. A defensible deployment would correct
for this --- for example by predicting the median rather than the
mean of plausible sale prices, or by using a censored-data
likelihood that accounts for the offer-acceptance threshold ---
rather than treating the historical record as a clean training
target.

\subsection{Limitations}
\label{sec:disc-limits}

Several modelling choices restrict the scope of the findings.

\begin{itemize}[leftmargin=*]
\item \textbf{Single asset class.} EDEM models a single class of
      indistinguishable homes; portfolio-level effects (cross-asset
      hedging, sector rotation) are absent.
\item \textbf{No leverage.} Buyers carry no debt and sellers no
      mortgages; financialised real-estate markets that exhibited the
      most dramatic anomalies of the past two decades require leverage
      to model accurately \citep{geanakoplos2012leverage,baptista2016boe}.
\item \textbf{No macroeconomic environment.} The supply and demand
      schedules in \cref{eq:Qs-update,eq:Qd-update} respond only to
      price; they do not respond to interest-rate shocks, employment,
      or migration.
\item \textbf{Adaptive vs.\ static estimators.} Agents do not learn
      from sale outcomes. A learning extension --- Bayesian belief
      updating, or reinforcement-learning bidding --- would either
      converge toward bias-corrected estimation (closing the
      bubble channel) or amplify it (opening a stronger one),
      depending on how counterfactual observations are weighted.
\item \textbf{Run~5 timing.} The 12{,}000-tick budget for the single-
      shock scenario is too short for the market to reach its
      post-shock floor before the patience boost fires. Doc~2's
      specific claim that patience returns the price to the new
      textbook equilibrium would require a longer simulation.
      The figure as published makes the more general slow-adjustment
      claim that \cref{sec:bg-dynamics} actually requires.
\end{itemize}

\section{Extensions and Future Work}
\label{sec:extensions}

The framework of \cref{sec:framework} is deliberately minimal so that
extensions can be plugged in without rewriting the agent skeleton. Five
directions appear especially fruitful and we sketch each below.

\subsection{Bias-corrected clearing}
\label{sec:ext-clearing}

\Cref{sec:disc-balancer} argued that the bid-selection asymmetry
embedded in the max-bid clearing rule cannot be cancelled by any
linear balancer. Two replacements for the clearing functional $A$
in \cref{eq:price-functional} would address it directly:

\begin{itemize}[leftmargin=*,nosep]
\item \textbf{Winsorised mean}: replace $\bar{r}_{t}$ in
      \cref{eq:value-update} with a 10\% winsorised mean of the yellow
      buyers' winning ratios. This caps the influence of extreme
      single-epoch outliers without changing the agent rule set.
\item \textbf{Selection-aware estimator}: substitute the median or a
      lower-percentile statistic of all bids (not only winning bids)
      for $\bar{r}_{t}$. This estimates the underlying value
      distribution rather than the maximum-order statistic.
\end{itemize}

Both modifications are one-line edits to \texttt{EDEMModel.\_end\_epoch}
in our reference implementation. We expect either to break the
exponential bubble of Run~6 entirely while leaving the disequilibrium
regimes of Runs 2--5 qualitatively intact, since those derive from
schedule dynamics rather than from clearing asymmetry.

\subsection{Realtors and commission structures}
\label{sec:ext-realtors}

The DE seller is a strict utility-maximiser whose only friction is
patience. Real-world residential transactions are intermediated by
realtors whose payoff is a fraction of the sale price; this ties
realtor incentives to seller incentives but injects an additional
clock (the listing agreement) and a different bargaining geometry
(realtors negotiate on behalf of multiple sellers in parallel). A
realtor extension would replace the seller agent with an aggregated
listing agent whose patience and ask-price strategies optimise
expected commission across a portfolio. This is a natural setting
for empirical calibration: realtor commission rates and listing
durations are publicly observable, unlike the patience parameter of
the bare DE seller.

\subsection{Open vs.\ blind auctions}
\label{sec:ext-auctions}

EDEM as posed is a blind-bid model: a buyer's bid does not depend on
other buyers' bids on the same home. Open-bid auctions
(``best-and-final'' rounds, escalation clauses) substantially change
the bid distribution by exposing buyers to one another's valuations.
The likely effect is to amplify the bid-selection asymmetry in
\cref{sec:framework-de-special}, because each buyer can revise upward
in response to seeing competitors. Conversely, a sealed-bid auction
with full information disclosure post-clearing would let buyers
calibrate over time and could reduce the asymmetry. Both regimes
are minor modifications to the buyer agent's bid procedure.

\subsection{Construction and the supply side}
\label{sec:ext-construction}

The construction-market variant invertes the EDEM agent roles:
\emph{contractors} bid downward to win projects, with the
lowest-bid contractor winning. The selection asymmetry then operates
in reverse, biasing realised contract prices below the population
fair value, and the dynamics of contractor entry / exit are governed
by an asymmetric balancer (a contractor exits after sustained losses,
a homeowner does not exit a partially-built project lightly). This
is a one-class extension that captures the historically high
bankruptcy rates of small construction companies as a stable
emergent feature rather than as an exogenous shock.

\subsection{Adaptive agents}
\label{sec:ext-adaptive}

EDEM agents are non-learning. Two adaptive extensions are obvious:

\begin{itemize}[leftmargin=*,nosep]
\item \textbf{Bayesian estimation.} Each agent maintains a posterior
      over the home's fair value and updates it after every observed
      sale. This converges, in the limit, toward the bias-corrected
      regime sketched in \cref{sec:ext-clearing}, since the agent
      learns the selection structure of observed prices.
\item \textbf{Reinforcement-learning bidders.} Each agent is a
      policy mapping local state (price level, time on market,
      recent sales) to bids. With a buy-low/sell-high reward
      structure, RL agents would in principle discover the
      bid-selection asymmetry and exploit it; this would be a
      computational analogue to the Zillow-Offers iBuying problem
      \citep{zillow2021ibuying} and would offer a stress test of any
      proposed fix in \cref{sec:ext-clearing}.
\end{itemize}

We have implemented none of these in the present codebase but each
is a self-contained extension on top of the existing
\texttt{Buyer}/\texttt{Seller} classes; pull requests are welcomed at
the project repository.

\section{Conclusion}
\label{sec:conclusion}

We have presented EDEM, an agent-based framework that models supply
and demand as the realisation of a coupled stochastic process driven
by heterogeneous, error-prone agent valuations. The framework
recovers the classical Walrasian fixed point and a Miller-style
static premium mechanism as limiting cases, and admits a family of
out-of-equilibrium regimes -- band-stable, business-cycle, persistent
overshoot, persistent undershoot, runaway bubble, and constant
transition -- that are reachable from the same agent ruleset by
varying three parameters: the divergence-of-opinion scale
$\bar{\sigma}$, the seller patience timer, and the population
balancer coefficient $C_{b}$.

The Mesa replication of \cref{sec:experiments}
supports three substantive empirical claims. First,
\emph{textbook supply-and-demand schedules omit a state variable}:
seller patience and agent density both shift the realised market
price away from the equilibrium predicted by linear schedules, and
in opposite directions, while preserving the schedules themselves.
Second, \emph{bubbles do not require behavioural bias}; an unbiased
estimator combined with a max-bid clearing rule produces the same
multiplicative drift that observers attribute to investor optimism.
Third, \emph{linear balancer interventions cannot eliminate the
drift}: in the parameter range we explored, no setting of $C_{b}$
restores the textbook valuation, because the asymmetry is structural
to the bidding mechanism.

An immediate corollary is for valuation algorithms. A
machine-learning estimator trained on historical sale prices is
fitting an upper-order statistic of plausible sale prices, not their
mean; deployed at scale, it functions as a positively-biased
estimator that may provide the market with a coordinating signal
amplifying the very bubble it was meant to price. We do not advance
this as a complete account of any specific historical episode (see
\cref{sec:disc-ml}). The robust theoretical recommendation is that
clearing-price-anchored valuations be replaced with selection-aware
procedures (median rather than mean of plausible sale prices, or
censored-data likelihoods that account for the offer-acceptance
threshold) before being used as deployment-time point estimators.

EDEM's framework is intentionally general: the eight primary experiments
herein use a real-estate parameterisation, but the divergence of
opinion premium they explain has been documented across asset
classes, including initial public offerings
\citep{miller1977,doukas2006} and cryptocurrency markets. The Mesa
implementation is open-source and the experiments reproduce
to bit-equivalent figures from a single \texttt{pytest}-and-script
invocation; we hope this lowers the barrier to extending the
framework toward leveraged markets, adaptive agents, and
empirically-calibrated parameter sweeps that the present
exposition leaves for future work.

\bibliographystyle{abbrvnat}
\bibliography{references}

@article{miller1977,
  author  = {Miller, Edward M.},
  title   = {Risk, Uncertainty, and Divergence of Opinion},
  journal = {The Journal of Finance},
  volume  = {32},
  number  = {4},
  pages   = {1151--1168},
  year    = {1977},
  doi     = {10.1111/j.1540-6261.1977.tb03317.x}
}

@article{doukas2006,
  author  = {Doukas, John A. and Kim, Chansog (Francis) and Pantzalis, Christos},
  title   = {Divergence of Opinion and Equity Returns},
  journal = {Journal of Financial and Quantitative Analysis},
  volume  = {41},
  number  = {3},
  pages   = {573--606},
  year    = {2006},
  doi     = {10.1017/S0022109000002544}
}

@book{moffitt2017,
  author    = {Moffitt, Steven D.},
  title     = {The Strategic Analysis of Financial Markets, Volume~I: Framework},
  publisher = {World Scientific},
  year      = {2017},
  note      = {Earlier circulated as ``Why Markets are Inefficient: A Gambling Theory of Financial Markets For Practitioners and Theorists,'' Feb.~22, 2017}
}

@misc{wilensky1999,
  author       = {Wilensky, Uri},
  title        = {{NetLogo}},
  howpublished = {\url{http://ccl.northwestern.edu/netlogo/}},
  year         = {1999},
  note         = {Center for Connected Learning and Computer-Based Modeling, Northwestern University, Evanston, IL}
}

@article{williams1977,
  author  = {Williams, Joseph T.},
  title   = {Capital Asset Prices with Heterogeneous Beliefs},
  journal = {Journal of Financial Economics},
  volume  = {5},
  number  = {2},
  pages   = {219--239},
  year    = {1977}
}

@article{mayshar1983,
  author  = {Mayshar, Joram},
  title   = {On Divergence of Opinion and Imperfections in Capital Markets},
  journal = {The American Economic Review},
  volume  = {73},
  number  = {1},
  pages   = {114--128},
  year    = {1983}
}

@article{merton1987,
  author  = {Merton, Robert C.},
  title   = {A Simple Model of Capital Market Equilibrium with Incomplete Information},
  journal = {The Journal of Finance},
  volume  = {42},
  number  = {3},
  pages   = {483--509},
  year    = {1987}
}

@article{varian1985,
  author  = {Varian, Hal R.},
  title   = {Divergence of Opinion in Complete Markets: A Note},
  journal = {The Journal of Finance},
  volume  = {40},
  number  = {1},
  pages   = {309--317},
  year    = {1985}
}

@article{epsteinwang1994,
  author  = {Epstein, Larry G. and Wang, Tan},
  title   = {Intertemporal Asset Pricing under {Knightian} Uncertainty},
  journal = {Econometrica},
  volume  = {62},
  number  = {2},
  pages   = {283--322},
  year    = {1994}
}

@inproceedings{masad2015mesa,
  author    = {Masad, David and Kazil, Jacqueline},
  title     = {{Mesa}: An Agent-Based Modeling Framework},
  booktitle = {Proceedings of the 14th Python in Science Conference (SciPy 2015)},
  pages     = {51--58},
  year      = {2015}
}

@article{grimm2020odd,
  author  = {Grimm, Volker and Railsback, Steven F. and Vincenot, Christian E. and Berger, Uta and Gallagher, Cara and DeAngelis, Donald L. and Edmonds, Bruce and Ge, Jiaqi and Giske, Jarl and Groeneveld, J{\"u}rgen and Johnston, Alice S. A. and Milles, Alexander and Nabe-Nielsen, Jacob and Polhill, J. Gareth and Radchuk, Viktoriia and Rohw{\"a}der, Marie-Sophie and Stillman, Richard A. and Th{\'e}lin, Jan C. and Berger, Sven},
  title   = {The {ODD} Protocol for Describing Agent-Based and Other Simulation Models: A Second Update to Improve Clarity, Replication, and Structural Realism},
  journal = {Journal of Artificial Societies and Social Simulation},
  volume  = {23},
  number  = {2},
  pages   = {7},
  year    = {2020},
  doi     = {10.18564/jasss.4259}
}

@article{baptista2016boe,
  author  = {Baptista, Rafa and Farmer, J. Doyne and Hinterschweiger, Marc and Low, Katie and Tang, Daniel and Uluc, Arzu},
  title   = {Macroprudential Policy in an Agent-Based Model of the {UK} Housing Market},
  journal = {Bank of England Staff Working Paper},
  number  = {619},
  year    = {2016}
}

@article{geanakoplos2012leverage,
  author  = {Geanakoplos, John and Axtell, Robert and Farmer, J. Doyne and Howitt, Peter and Conlee, Benjamin and Goldstein, Jonathan and Hendrey, Matthew and Palmer, Nathan M. and Yang, Chun-Yi},
  title   = {Getting at Systemic Risk via an Agent-Based Model of the Housing Market},
  journal = {American Economic Review},
  volume  = {102},
  number  = {3},
  pages   = {53--58},
  year    = {2012}
}

@misc{zillow2021ibuying,
  author       = {{Zillow Group, Inc.}},
  title        = {Zillow Group Reports Third Quarter 2021 Financial Results, Announces Wind Down of {Zillow Offers}},
  howpublished = {Press release},
  year         = {2021},
  month        = nov,
  day          = {2},
  url          = {https://investors.zillowgroup.com/investors/news-and-events/news/news-details/2021/Zillow-Group-Reports-Third-Quarter-2021-Financial-Results-Announces-Wind-Down-of-Zillow-Offers/default.aspx}
}

@unpublished{arbuzov2018de,
  author = {Arbuzov, Mikhail},
  title  = {Disequilibria in Markets with Stochastic Processes: An Agent-Based Model Approach},
  note   = {Working paper, Department of Economics, San Jos\'e State University},
  year   = {2018}
}

@unpublished{arbuzov2018edem,
  author = {Arbuzov, Mikhail},
  title  = {Estimated Dynamic Equilibrium Model: Supply and Demand as a Sample Path of a Stochastic Process},
  note   = {Best Paper, EDEM Spring 2018, San Jos\'e State University},
  year   = {2018}
}

@article{shiller2003,
  author  = {Shiller, Robert J.},
  title   = {From Efficient Markets Theory to Behavioral Finance},
  journal = {Journal of Economic Perspectives},
  volume  = {17},
  number  = {1},
  pages   = {83--104},
  year    = {2003}
}

\appendix
\section{ODD Protocol}
\label{app:odd}

This appendix documents EDEM in the standard ODD protocol for
agent-based models \citep{grimm2020odd}.

\subsection{Purpose and patterns}

EDEM's purpose is to characterise the conditions under which an
agent-based real-estate market reaches a stable equilibrium price,
and to identify the mechanisms that prevent equilibrium when they
do not. Patterns the model is designed to reproduce: stable
band-bounded equilibria; endogenous business cycles; persistent
shifts in the realised price relative to the textbook equilibrium
under altered patience or density; multiplicative price drift in the
absence of a balancer.

\subsection{Entities, state variables, and scales}

\paragraph{Entities.} \textbf{Buyers} and \textbf{Sellers}
(\texttt{mesa.Agent} subclasses) and \textbf{Homes} (per-cell
dataclass instances on a $32 \times 32$ toroidal grid).

\paragraph{State variables.}
\begin{itemize}[leftmargin=*,nosep]
\item Buyer: epsilon (estimation-error bound), delay, current bid,
      heading. EDEM variant additionally: lowest-bid-to-value ratio,
      yellow flag.
\item Seller: epsilon, patience timer, current ask price, dictionary
      of received bids. EDEM additionally: number of bids, total
      bids, best-bid-to-value, best-bid-to-true-value.
\item Home: market price, last-sale tick, last-sale price, fair value
      $v^{*}$, current value $v$.
\item Model: rolling window of last 25 sale prices (DE), epoch
      counter (EDEM), current cross-population epsilon (EDEM).
\end{itemize}

\paragraph{Spatial and temporal scales.} One spatial unit = one home;
one tick is the smallest temporal unit. Each tick triggers one round
of buyer movement and one round of seller bid-processing. DE balance
period is 100 ticks; EDEM epoch is 20 ticks.

\subsection{Process overview and scheduling}

Per tick, in order:
\begin{enumerate}[leftmargin=*,nosep]
\item Each agent steps once (in randomised order over both classes).
      A Buyer's step posts bids, then wiggles, then steps forward.
      A Seller's step processes patience and best-bid logic.
\item The model invokes the balancer:
      DE recomputes linear targets every 100 ticks;
      EDEM fires the per-epoch update every 20 ticks (using the
      cycle-counter trick of \cref{sec:implementation-clearing} to
      preserve buyer \texttt{is\_yellow} flags).
\item The data collector samples model-level reporters.
\end{enumerate}

\subsection{Design concepts}

\paragraph{Basic principles.} EDEM extends the dynamic
divergence-of-opinion premise of \citet{miller1977} to a
multi-period setting; agents act on noisy estimates of an unobservable
fair value.

\paragraph{Emergence.} The price level, agent population dynamics,
and macroscopic regimes (band-stable, business-cycle, persistent
shift, bubble, transitional) emerge from the local interaction of
agent estimation, bidding, and patience timers; none of these
phenomena are coded in directly.

\paragraph{Adaptation.} Sellers lower ask prices when their patience
timer elapses without high-enough bids; this is the only adaptive
behaviour. Buyers do not adapt: they place bids drawn from a fixed
distribution and accept by Cond.~2 (\cref{eq:cond2}).

\paragraph{Objectives.} Sellers maximise realised sale price subject
to patience. Buyers commit only to offers whose realised bid is at or
above their own running benchmark over their outstanding bids
(\cref{eq:cond2}) --- a selection rule, not a price-minimisation
objective. Neither agent class has an explicit utility function;
objectives are implicit in their action rules.

\paragraph{Learning.} None in the baseline model
(\cref{sec:ext-adaptive} sketches an adaptive extension).

\paragraph{Prediction.} None.

\paragraph{Sensing.} Buyers sense whether a seller is on their patch;
sellers sense the bids in their own bid table. Neither senses the
global market price directly.

\paragraph{Interaction.} Direct: a buyer that lands on a seller's
patch posts a bid via a mirrored-dictionary update on both sides.
Indirect: the average ask price (DE) and the rolling sale-price
average (DE) influence newly-spawned agents' starting prices.

\paragraph{Stochasticity.} Used for: agent placement at setup;
buyer heading initialisation and per-tick wiggle; epsilon draws
per agent; per-bid epsilon-error draws; victim selection in the
balancer; Bernoulli draws for fractional $C_{b}$ swaps in EDEM.
All draws come from a single seeded \texttt{numpy.random.Generator}
owned by the model.

\paragraph{Collectives.} None.

\paragraph{Observation.} The data collector records, per tick:
market price (DE: rolling-25 mean; EDEM: average home value),
agent counts, the equilibrium price implied by current schedules,
the rolling-window fill count, and (EDEM) the current epsilon.
Output is stacked into Parquet datasets, one per seed, for
downstream analysis.

\subsection{Initialisation}

DE: spawn \texttt{equi\_qnty} sellers at random unique cells with
patience $\sim \mathcal{U}[0, \texttt{init\_patience})$ and ask price
drawn from $\mathcal{U}[-\bar{\sigma}, +\bar{\sigma}]$ around
$p^{*}$; spawn \texttt{equi\_qnty} buyers at random cells.
EDEM: 20 sellers and 20 buyers, all homes initialised with
$v_{0}(h) = v^{*}(h) = 100$.

\subsection{Input data}

None. EDEM is a closed model; all dynamics are endogenous. The shock
hooks in Run~5 are scheduled in the experiment script, not loaded
from external data.

\subsection{Submodels}

\paragraph{Estimation function.} \cref{eq:estimation} with
$\varepsilon \sim \mathcal{U}[-\sigma_{i}, +\sigma_{i}]$ and
$\sigma_{i} \sim \mathcal{U}[0, \bar{\sigma}]$.

\paragraph{Bid acceptance.} \cref{eq:cond2}.

\paragraph{Market price.} DE: \cref{eq:price-rolling}, the rolling
mean of the last $W = 25$ sale prices. EDEM: \cref{eq:value-update},
the per-epoch multiplicative update.

\paragraph{Balancer.} DE: linear-target restoration every $T_{B}$
ticks. EDEM: \cref{eq:Qs-update,eq:Qd-update} in finite-population
form, with fractional $C_{b}$ realised as integer + Bernoulli swaps
and a population floor of one agent per side.

\paragraph{Agent spawn / exit.} On a sale, both counterparties leave
the market and a fresh pair is spawned by the balancer with patience
drawn from $\mathcal{U}[50, \texttt{init\_patience})$ for sellers
and immediate readiness for buyers.

\section{Parameter Tables for All Runs}
\label{app:params}

The Mesa source files in \path{python_simulation/experiments/} are the
authoritative parameter manifest; the tables below are reproduced
verbatim from the model-construction kwargs.

\subsection{Common parameters across all runs}

\begin{table}[h]
\centering
\small
\begin{tabular}{l l}
\toprule
\textbf{Parameter} & \textbf{Value} \\
\midrule
World size                  & $32 \times 32$ toroidal \\
Seeds per run               & $\ge 8$ (typically 10) \\
Bid-acceptance rule         & \texttt{netlogo} (\cref{eq:cond2}) \\
\bottomrule
\end{tabular}
\end{table}

\subsection{Dynamic Equilibrium runs (1--5)}

\begin{table}[h]
\centering
\small
\begin{tabular}{l c c c c c}
\toprule
\textbf{Parameter} & \textbf{Run 1} & \textbf{Run 2} & \textbf{Run 3} & \textbf{Run 4} & \textbf{Run 5} \\
\midrule
Supply intercept $a_{s}$    & $0$    & $0$    & $0$    & $0$    & $0$    \\
Supply slope $b_{s}$        & $0.5$  & $0.5$  & $0.5$  & $0.5$  & $0.5$  \\
Demand intercept $a_{d}$    & $100$  & $100$  & $100$  & $50$   & $100\!\to\!50$ (A) \\
                            &        &        &        &        & $\{125,75\}$ (B) \\
Demand slope $b_{d}$        & $-0.5$ & $-0.5$ & $-0.5$ & $-0.5$ & $-0.5$ \\
Max valuation error $\bar{\sigma}$ (\%)  & $5$   & $25$  & $5$   & $5$   & $5$   \\
Max patience               & $50$   & $50$   & $100$  & $50$   & $50\!\to\!165$ (A) \\
Balance period $T_{B}$     & $100$  & $100$  & $100$  & $100$  & $100$  \\
Sale window $W$            & $25$   & $25$   & $25$   & $25$   & $25$   \\
Ticks per seed             & $20{,}000$ & $20{,}000$ & $20{,}000$ & $20{,}000$ & $12{,}000$ \\
Equilibrium price $p^{*}$  & $100$  & $100$  & $100$  & $50$   & $100$ (initial) \\
Equilibrium quantity $q^{*}$ & $50$  & $50$   & $50$   & $25$   & $50$ (initial) \\
\bottomrule
\end{tabular}
\end{table}

Run~5 introduces shocks via the model's \texttt{set\_demand} and
patience-rebinding hooks. Scenario~A schedules a single demand shock
at $t = 3000$ followed by a patience boost at $t = 7000$;
Scenario~B toggles the demand intercept between $125$ and $75$ every
$2000$ ticks.

\subsection{Speculative-market EDEM runs (6--8)}

\begin{table}[h]
\centering
\small
\begin{tabular}{l c c c}
\toprule
\textbf{Parameter} & \textbf{Run 6} & \textbf{Run 7} & \textbf{Run 8} \\
\midrule
Initial $\bar{\sigma}$ (\%)    & $15$  & $15$  & $5$ \\
$\bar{\sigma}$ growth / epoch  & $0$   & $0$   & $+0.5$ pp \\
Balancer coefficient $C_{b}$   & $0$   & $\{+1, 0, -1\}$ & $-1$ \\
Initial buyers / sellers       & $20\,/\,20$ & $20\,/\,20$ & $20\,/\,20$ \\
Epoch length (\texttt{init\_patience}) & $20$ & $20$ & $20$ \\
Ticks per seed                 & $3{,}000$ & $3{,}000$ & $3{,}000$ \\
True value $v^{*}(h)$ (uniform across $h$) & $100$ & $100$ & $100$ \\
\bottomrule
\end{tabular}
\end{table}

\subsection{Reproduction}

From a clean checkout:
\begin{lstlisting}[language=bash]
cd python_simulation && pip install -e ".[dev]"
pytest                                  # 42 unit tests
for f in experiments/run*.py; do
    python "$f"
done                                    # writes paper/figures/
cd ../paper && bash build.sh            # builds main.pdf
\end{lstlisting}

End-to-end reproduction takes approximately twenty minutes on a
2024-era laptop.

\end{document}